 \documentclass[preprint, 12pt]{elsarticle}

\usepackage{amssymb}
\usepackage{bm}
\usepackage[colorlinks=true, linkcolor=blue, anchorcolor=blue,citecolor=blue]{hyperref}
\usepackage{amsmath}
\usepackage{caption}
\usepackage{cleveref}
\usepackage{multirow}
\usepackage{color,soul}
\usepackage{url}
\usepackage[table]{xcolor}
\biboptions{compress}
\newcommand{\degree}{^\circ}

\begin{document}
\sethlcolor{yellow}
\setstcolor{red}
\soulregister\cite7 
\soulregister\ref7

\begin{frontmatter}

\title{An Edge-based Interface Tracking (EBIT) Method for Multiphase Flows with Phase Change}

\author{Tian Long\fnref{label1,label2}}
\ead{tian.long@dalembert.upmc.fr}

\author{Jieyun Pan\corref{cor1}\fnref{label2}}
\ead{yi.pan@sorbonne-universite.fr}
\cortext[cor1]{Corresponding author}

\author{Stéphane Zaleski\fnref{label2,label3}}
\ead{stephane.zaleski@sorbonne-universite.fr}

\address[label1]{School of Aeronautics, Northwestern Polytechnical University, Xi’an, 710072, PR China}
\address[label2]{Sorbonne Université and CNRS, Institut Jean Le Rond d’Alembert UMR 7190, F-75005 Paris, France}
\address[label3]{Institut Universitaire de France, Paris, France}

\begin{abstract}
In this paper, the Edge-Based Interface Tracking (EBIT) method is extended to simulate multiphase flows with phase change. As a novel Front-Tracking method, the EBIT method binds interfacial markers to the Eulerian grid to achieve automatic parallelization. To include phase change effects, the energy equation for each phase is solved, with the temperature boundary condition at the interface sharply imposed. When using collocated grids, the cell-centered velocity is approximately projected. This will lead to unphysical oscillations in the presence of phase change, as the velocity will be discontinuous across the interface. It is demonstrated that this issue can be addressed by using the ghost fluid method, in which the ghost velocity is set according to the jump condition, thereby removing the discontinuity. A series of benchmark tests are performed to validate the present method. It is shown that the numerical results agree very well with the theoretical solutions and the experimental results.  
\end{abstract}

\begin{keyword}
   Front-Tracking\sep Multiphase flows\sep Phase change\sep Ghost fluid method \sep Collocated grid
    
\end{keyword}

\end{frontmatter}


\section{Introduction}
\label{sec1}
Multiphase flows with phase change play a critical role in various industrial applications, including combustion engines \cite{mayer1996propellant}, electronic cooling systems \cite{liu2020assessment}, and nuclear reactors \cite{seong2022effect}. A deeper understanding of the associated mass and heat transfer processes is essential for advancing these technologies \cite{dhir2006mechanistic,kim2021effects}. However, the multiscale nature of phase change flows often makes experimental studies formidable. One example is nucleate boiling in the micro-layer regime \cite{son1999dynamics, burevs2021modelling}, where a thin film, only a few microns thick, forms between the bubble and the heated wall. In this micro-layer, strong heat transfer occurs and contributes significantly to the overall heat extraction. Conducting quantitative measurements within this micro-layer proves to be a challenging task \cite{zupanvcivc2022wall}. Therefore, numerical simulation, with its ability to detail physical processes at microscopic levels \cite{urbano2018direct, urbano2019direct}, has become a powerful tool for studying phase change flows and has gained increasing interest in academia. In recent years, various computational methods have been developed to simulate multiphase flows with phase change \cite{ kharangate2017review}. Generally, based on the description \textcolor{blue}{of} the phase interface, these methods can be classified into Front-Capturing and Front-Tracking methods \cite{tryggvason2011direct}. In Front-Capturing methods, the interface movement is implicitly captured via the integration of a tracer function over time, for example, the Heaviside function in Volume-of-Fluid (VOF) methods \cite{hirt1981volume,brackbill1992continuum}, the signed distance function in Level-Set (LS) methods \cite{osher1988fronts, osher2001level}, and the phase-field equation in Phase-Field (PF) methods \cite{ding2007diffuse,zhao2023interaction}. These methods can automatically handle topology changes and allow for efficient parallelization, making them popular for simulating phase change flows. For a detailed implementation of phase change models based on these methods, readers are referred to Refs. \cite{malan2021geometric, boyd2023consistent, cipriano2023multicomponent} for VOF, Refs. \cite{tanguy2014benchmarks,dhruv2019formulation, long2023fully} for LS, and Refs. \cite{ruyer2006phase,satenova2021simulation} for PF.

Compared with Front-Capturing methods, Front-Tracking methods \cite{glimm1985computational,unverdi1992front} explicitly track interface motion by advecting connected marker points. Their capability to track and control topology makes them superior for complex multiscale problems \cite{chen2022characterizing}. Simulations of phase change flows using Front-Tracking can be traced back to the work of Juric and Tryggvason \cite{juric1998computations}, which investigates 2D film boiling. In this work, an iterative procedure is employed to impose the correct temperature boundary condition at the interface. Esmaeeli and Tryggvason \cite{esmaeeli2004computations1, esmaeeli2004computations2} later eliminate this iterative process in their improved algorithm, which is subsequently used to study multi-mode film boiling on a horizontal surface. Furthermore, dendrite solidification \cite{al2002numerical, al2004numerical} and film boiling from multiple horizontal cylinders \cite{esmaeeli2004front} are studied using Front-Tracking and the immersed boundary method \cite{fadlun2000combined}. The three-phase droplet icing problem, including tri-junction and volume change, is computed in Refs. \cite{vu2013front,vu2015numerical}, where three types of fronts are tracked. Additionally, Irfan and Muradoglu \cite{irfan2017front} extend Front-Tracking to droplet evaporation driven by species gradients. All these phase change models are based on the Front-Tracking method of Unverdi and Tryggvason \cite{unverdi1992front}, which requires manual handling of topology changes. In contrast, the Level Contour Reconstruction Method (LCRM) \cite{shin2002modeling} and the Local Front Reconstruction Method (LFRM) \cite{shin2011local} eliminate the need to store logical connections between neighboring surface elements, thus enabling robust simulations of 3D flows exhibiting topology changes with Front-Tracking. They have been successfully applied to film boiling \cite{shin2002modeling}, nucleate boiling \cite{shin2005direct}, and rising bubbles with phase change \cite{shin2016numerical,rajkotwala2019critical}.

In this work, using the free software Basilisk \cite{popinet2009accurate,popinet2015quadtree}, we aim to extend a novel Front-Tracking method, the Edge-Based Interface Tracking (EBIT) method \cite{chirco2023edge,pan2023edge}, to simulate multiphase flows with phase change. The EBIT method, without storing connectivity information, binds Lagrangian markers to the Eulerian grid to achieve automatic parallelization. Its capability and accuracy for multiphase flow without phase change have been validated through typical benchmark problems \cite{pan2023edge}. To extend the EBIT method to phase change flows, we solve additional energy equations with the temperature boundary condition sharply imposed at the interface. The mass flux is computed from the heat fluxes at the interface and is used to solve the adjusted mass and momentum equations. In the presence of phase change, the velocity is discontinuous across the interface. As we employ a collocated grid, on which the cell-centered velocity is approximately projected, unphysical oscillations will be introduced near the interface. To suppress unphysical oscillations, we adopt the ghost fluid method and solve two separate velocity fields. For each phase, the ghost velocity is determined by the real velocity of the other phase and the jump condition due to phase change, thus removing the discontinuity at the interface. The remainder of this paper is organized as follows: In Section 2, we briefly review the EBIT method for completeness. Then, we introduce how to solve the energy equation and how to incorporate the mass flux into the solution of the mass and momentum equations via the ghost fluid method. Subsequently, in Section 3, we validate the present method with several benchmark tests, followed by concluding remarks in Section 4.

\section{Numerical Method}
\label{sec2}
\subsection{The EBIT method}
\begin{figure}[htbp]
    \centering
    \includegraphics[width=1.0\textwidth]{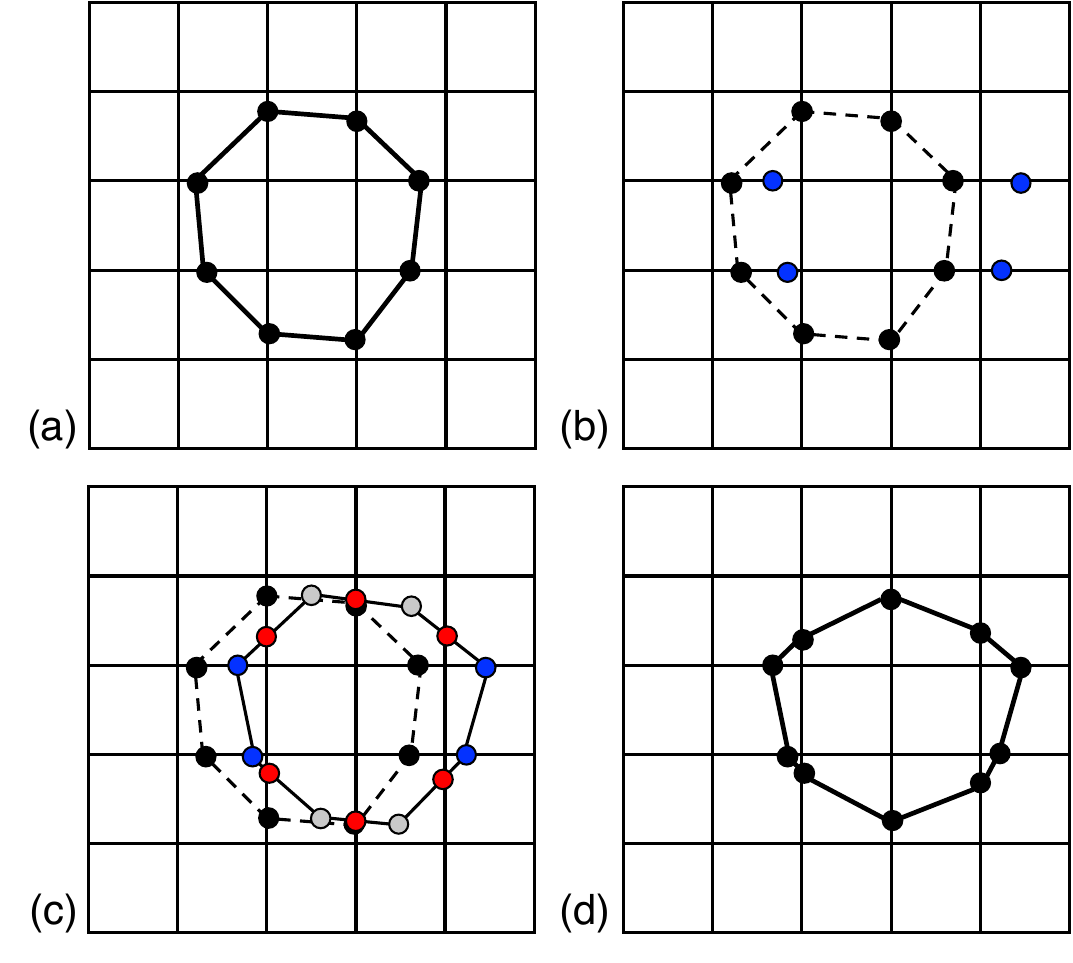}
    \caption{Illustration of the 1D advection of the EBIT method along the x-direction: (a) Initial interface. (b) Aligned markers after advection (blue points). (c) Unaligned markers after advection (grey points) and the newly created markers intersecting with grid lines (red points). Note that the grey points will be discarded. (d) Interface after 1D advection. A marker is defined as aligned or unaligned according to whether the grid line it is located on is aligned with the current advection direction.}
    \label{Fig:1D_advecttion_illustration}
\end{figure}

In this paper, the EBIT method \cite{chirco2023edge, pan2023edge} is employed for the interface tracking, in which the interfacial markers are restricted to be moved along the grid lines during the interface advection. Note that only one marker is allowed per cell edge in the current work. Let $\mathbf{x}_i$ denote the position for a given marker, its motion is described by
\begin{equation}
\frac{d \mathbf{x}_i}{dt} = \mathbf{u}_{\Gamma, i},
\label{Eq:marker_motion}
\end{equation}
where the interfacial velocity $\mathbf{u}_{\Gamma}$ is obtained from the flow field. By employing the first-order explicit Euler method, the discretization of Eq. (\ref{Eq:marker_motion}) reads
\begin{equation}
\mathbf{x}_i^{n+1} = \mathbf{x}_i^{n} + \mathbf{u}_{\Gamma, i}^n \Delta t,
\label{Eq:discrete_marker_motion}
\end{equation}
with the superscripts $n$ and $n+1$ representing successive discrete time instants. To solve the multi-dimensional interface advection, we adopt the dimension-splitting method \cite{chirco2023edge} and advect the interface along each dimension successively. In each 1D advection, a marker is defined as either an aligned or unaligned marker, depending on whether it is located on the grid line that is aligned with the current advection direction. The example of 1D advection along the x-direction is given in Fig. \ref{Fig:1D_advecttion_illustration}. As shown in Fig. \ref{Fig:1D_advecttion_illustration}(b), the new positions of aligned markers are directly obtained by solving Eq. (\ref{Eq:discrete_marker_motion}). In contrast, an unaligned marker will first be advected by solving Eq. (\ref{Eq:discrete_marker_motion}) (the grey points in Fig. \ref{Fig:1D_advecttion_illustration}(c)), followed by fitting a circle through the surrounding marker points to find the intersection with the unaligned grid line (the red points in Fig. \ref{Fig:1D_advecttion_illustration}(c)). Note that the intermediate unaligned markers that leave the grid lines are finally discarded. 

After advection, the connectivity of markers is reconstructed with the help of the Color Vertex method \cite{singh2007three}, where a color field is adopted to indicate the corresponding fluid phase. As shown in Ref. \cite{pan2023edge}, for the 2D situation, five color vertices (four in the corners and one in the cell center) are sufficient to determine the topological configuration and reconstruct the interface segments without ambiguity. Furthermore, with the Color Vertex method, topology changes are handled automatically. For more details about the EBIT method, readers are referred to Refs. \cite{chirco2023edge,pan2023edge}.

\subsection{Governing equations}
Consider a flow domain occupied by the liquid phase and the vapor phase, which are separated by an evolving interface $\Gamma (t)$. By assuming the two phases are both incompressible and monocomponent, the governing equations \cite{tanguy2014benchmarks} are 
\begin{align}
    \nabla \cdot \mathbf{u} &= S_{pc}, \label{Eq:mass_equation}\\
    \rho \left( \frac{\partial\mathbf{u}}{\partial t} + \mathbf{u} \cdot \nabla \mathbf{u}  \right) &= -\nabla p + \nabla \cdot \left( \mu ( \nabla \mathbf{u} + \nabla \mathbf{u}^T) \right) + \rho \mathbf{g} + \mathbf{f}_\sigma, \label{Eq:momentum_equation}\\
    \rho C_p \left( \frac{\partial T}{\partial t} + \mathbf{u} \cdot \nabla T\right) & = \nabla \cdot (\lambda \nabla T),\label{Eq:energy_equation}
\end{align}
where $\mathbf{u}$ is the fluid velocity, $\rho$ the density, $\mu$ the dynamic viscosity, $\mathbf{g}$ the gravity acceleration, $T$ the temperature, $C_p$ the heat capacity per unit mass at constant pressure, and $\lambda$ the thermal conductivity. The source term due to phase change $S_{pc}$ will be given later, and the surface tension term $\mathbf{f}_\sigma = \sigma \kappa \delta_s \mathbf{n}$ is computed by using a well-balanced Continuous Surface Force (CSF) method \cite{popinet2009accurate, popinet2018numerical}, where $\sigma$ is the surface tension coefficient, $\kappa$ the interface curvature, $\mathbf{n}$ the interface normal vector and $\delta_s$ the Dirac distribution function concentrated at the interface.

Across the interface, fluid properties will change, which can be calculated by
\begin{equation}
    \phi = \phi_{liq}H  + \phi_{vap}(1 - H) ,
    \label{Eq:property_average}
\end{equation}
where $\phi$ represents an arbitrary variable, such as density and viscosity. The subscripts $liq$ and $vap$ indicate the physical properties of liquid and vapor, respectively. The Heaviside function $H$ is defined as $1$ in the liquid region and $0$ in the vapor region. In numerical simulations, Eq. (\ref{Eq:property_average}) is typically computed by adopting the volume fraction \cite{tryggvason2011direct}, which is the integral of $H$ over a given cell. Note that, in this work, volume fractions are directly computed from the reconstructed interface given by the EBIT method. Furthermore, at the interface, the mass flux $\dot{m}$ in the presence of phase change is defined as
\begin{equation}
        \dot{m} = \rho_{liq} (\mathbf{u}_{liq} - \mathbf{u}_\Gamma )\cdot \mathbf{n} = \rho_{vap} (\mathbf{u}_{vap} - \mathbf{u}_\Gamma)\cdot \mathbf{n}. 
        \label{Eq:mass_flux}
\end{equation}
With the jump operator $[\phi] = \phi_{liq} - \phi_{vap}$, Eq. (\ref{Eq:mass_flux}) can be reformulated and leads to the velocity jump condition
\begin{equation}
        [\mathbf{u}] = \dot{m}\left[ \frac{1}{\rho} \right] \mathbf{n}, 
        \label{Eq:velocity_jump}
\end{equation}
and the jump condition for pressure is given by
\begin{equation}
   [p] = \sigma \kappa + 2\left[\mu \frac{\partial \mathbf{u}}{\partial n} \right] - \dot{m}^2 \left[ \frac{1}{\rho} \right].
   \label{Eq:jump_pressure}
\end{equation}
By considering the divergence-free condition in the bulk region of each phase and the velocity jump at the interface, the source term due to phase change $S_{pc}$ \cite{sato2013sharp} becomes 
\begin{equation}
   S_{pc} = \frac{\dot{m} S_\Gamma}{V}\left[ \frac{1}{\rho} \right],
   \label{Eq:pc_source}
\end{equation}
where $V$ is the volume of the computational cell and $S_{\Gamma}$ denotes the area of the interface within it.

\subsection{Solving the mass and momentum equations}
The present work utilizes the free software Basilisk \cite{popinet2009accurate,popinet2015quadtree} and employs a time-staggered approximate projection method to solve the incompressible Navier-Stokes equations. Spatial discretization is achieved using a collocated grid where all variables are stored at the cell center. The advection term is discretized using the Bell-Colella-Glaz (BCG) scheme \cite{bell1989second}, while the diffusion term is solved implicitly. In cases without phase change, the momentum equation Eq. (\ref{Eq:mass_equation}) is discretized as follows:
\begin{align}
&\rho_c^{n+\frac{1}{2}}\left(\frac{\mathbf{u}^*-\mathbf{u}^n}{\Delta t}+\mathbf{u}^{n+\frac{1}{2}} \cdot \nabla \mathbf{u}^{n+\frac{1}{2}}\right)_c=\nabla_c \cdot\left[\mu_f^{n+\frac{1}{2}}\left(\nabla \mathbf{u}+\nabla \mathbf{u}^T\right)^*\right]+\left[\left(\sigma \kappa \delta_s \mathbf{n}\right)^{n-\frac{1}{2}}-\nabla p^n\right]_{f \rightarrow c}, 
\label{Eq:discretization_1}\\
&\mathbf{u}_c^{* *}=\mathbf{u}_c^*-\frac{\Delta t}{\rho_c^{n+\frac{1}{2}}}\left[\left(\sigma \kappa \delta_s \mathbf{n}\right)^{n-\frac{1}{2}}-\nabla p^n\right]_{f \rightarrow c}, 
\label{Eq:discretization_2}\\
&\mathbf{u}_f^{*}=\mathbf{u}_{c \rightarrow f}^{* *} + \frac{\Delta t}{\rho_f^{n+\frac{1}{2}}}\left(\sigma \kappa \delta_s \mathbf{n}\right)^{n+\frac{1}{2}}, 
\label{Eq:discretization_3}\\
&\mathbf{u}_f^{n+1}=\mathbf{u}_f^{*} - \frac{\Delta t}{\rho_f^{n+\frac{1}{2}}}\nabla p^{n+1},
\label{Eq:discretization_4}\\
&\mathbf{u}_c^{n+1}=\boldsymbol{u}_c^{* *}+\frac{\Delta t}{\rho_c^{n+\frac{1}{2}}}\left[\left(\sigma \kappa \delta_s \boldsymbol{n}\right)^{n+\frac{1}{2}}-\nabla p^{n+1}\right]_{f \rightarrow c},
\label{Eq:discretization_5}
\end{align}
where the subscripts $c$ and $f$ denote the cell-centered variable and the face-centered variable, respectively. A second-order accurate scheme \cite{zhao2022boiling} is employed to interpolate the variables between the cell center and the cell face, which is denoted by the symbol $c \rightarrow f$ or $f \rightarrow c$. Note that the variable at different time steps is denoted by the superscripts $n-\frac{1}{2}$, $n+\frac{1}{2}$, $n$, and $n+1$, respectively. In Eq. (\ref{Eq:discretization_4}), the pressure at time $n+1$ is obtained by solving the Poisson equation,
\begin{equation}
\nabla_c \cdot\left(\frac{\Delta t}{\rho^{n+\frac{1}{2}}_f} \nabla p^{n+1}\right)=\nabla_c \cdot \mathbf{u}_f^*-\nabla_c \cdot \mathbf{u}^{n+1}_f,
     \label{Eq:Poisson_equation}
\end{equation}
with the mass conservation equation $\nabla_c \cdot \mathbf{u}_f^{n+1} = 0$ imposed. It can be seen that, though $\mathbf{u}_f$ is exactly projected onto a divergence-free velocity field, $\mathbf{u}_c$ is only approximately projected \cite{zhao2022boiling}, as it is interpolated using Eq. (\ref{Eq:discretization_5}). The use of the cell-centered velocity $\mathbf{u}_c$ increases the complexity of the algorithm. Nevertheless, it facilitates the implementation of  the implicit Crank-Nicolson discretization of the viscous terms on quad/octree adaptive meshes, which is formally second-order accurate and more efficient owing to its unconditional stability\cite{popinet2009accurate}. Although this approximate projection method works well for flows without phase change \cite{popinet2009accurate}. However, it may incur numerical instability near the interface in the presence of phase change \cite{zhao2022boiling}.

\begin{figure}[htbp]
    \centering
    \includegraphics[width=1.0\textwidth]{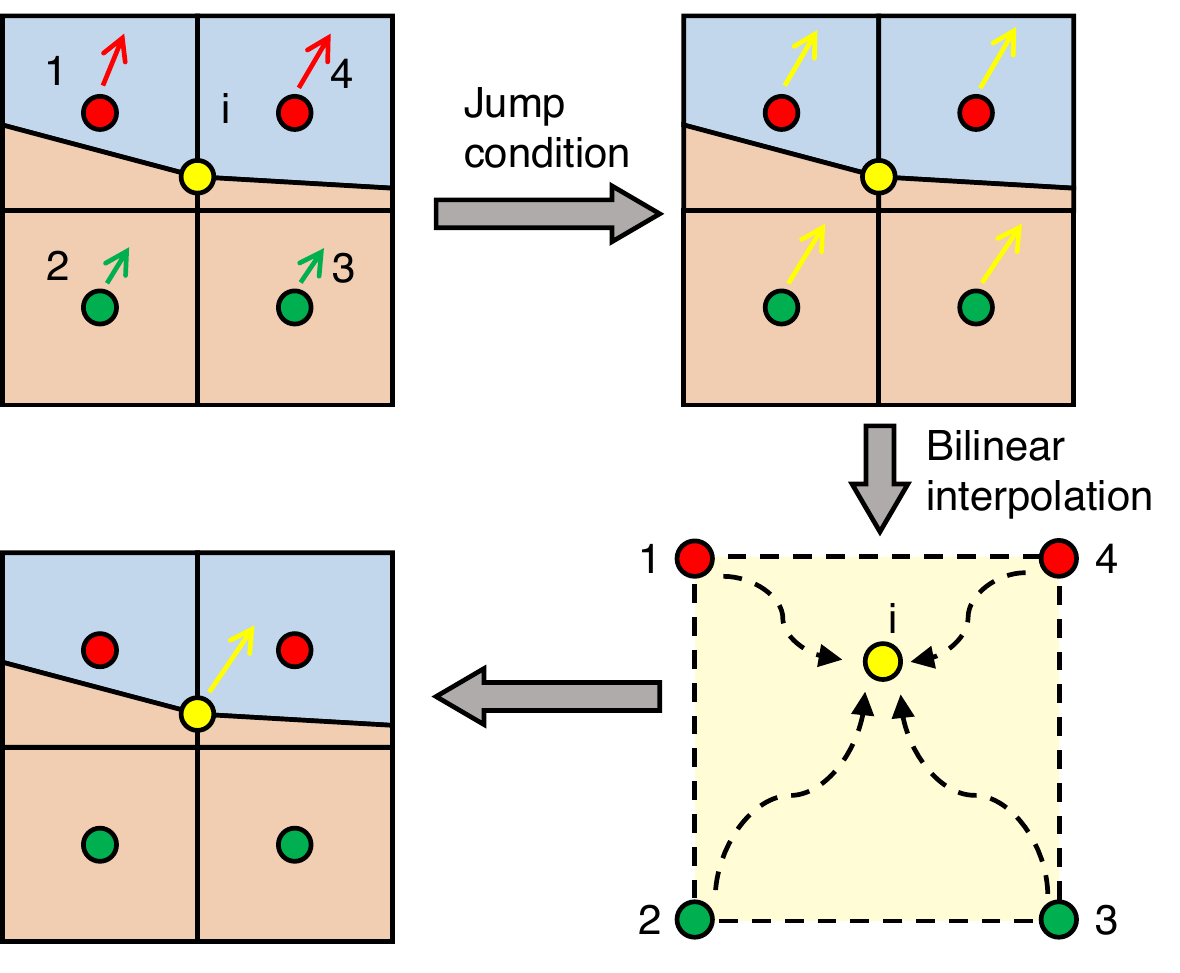}
    \caption{Schematic of the computation of interfacial velocity: The value at the i-th interfacial marker is derived from the nearest background grid cells 1-4. The green and red points represent the vapor and liquid cell centers, respectively, while the yellow point denotes the interfacial marker. The fluid velocities (indicated by the green and red arrows) are obtained by solving the momentum equation. These velocities are then converted to the interfacial velocity (denoted by the yellow arrow) using the jump condition (Eq. (\ref{Eq:jump_interfacial_vel})), and interpolated to the interfacial marker through bilinear interpolation.}
    \label{Fig:intefacial_vel_calculation}
\end{figure}

When phase change is considered, the mass flux is computed after solving the energy equation, which will be elaborated in the next section. To incorporate the mass flux resulting from phase change into the mass and momentum equations, there are mainly two types of methods: the one fluid method \cite{sato2013sharp} or the ghost fluid method \cite{tanguy2014benchmarks}. The implementation of the one fluid method is straightforward. During the solution of the Poisson equation, a source term $S_{pc}$, which is nonzero only in the interface cells (see Eq. (\ref{Eq:pc_source})), is added to the mass conservation equation: $\nabla_c \cdot \mathbf{u}_f^{n+1} = S_{pc}$. The singular source term may lead to numerical oscillations near the interface, especially for the cell-centered velocity $\mathbf{u}_c$ which is only approximately projected. To avoid such oscillations, Zhao et al. \cite{zhao2022boiling} proposed an exact projection method, solving an additional Poisson equation to ensure that $\mathbf{u}_c$ is exactly projected. In this work, we employ the more efficient ghost fluid method \cite{tanguy2014benchmarks}. By setting the ghost velocity, the singularity at the interface is removed. As will be shown in the numerical tests, numerical stability can be effectively improved by employing the ghost fluid method. With the jump condition Eq. (\ref{Eq:jump_pressure}), the ghost velocity is populated by
\begin{equation}
    \begin{aligned}
     \mathbf{u}_{vap}^{ghost} &= \mathbf{u}_{liq} - \dot{m}\left[ \frac{1}{\rho} \right] \mathbf{n} &\quad \text{if $f_C = 1$}, \\
    \mathbf{u}_{liq}^{ghost} &= \mathbf{u}_{vap} + \dot{m}\left[ \frac{1}{\rho} \right] \mathbf{n} &\quad \text{if $f_C = 0$}, \\ 
    \end{aligned}
    \label{Eq:ghost_velocity}
\end{equation}
where the color function $f_C$ is naturally provided by the Color Vertex technique used in the EBIT method, being $1$ in the liquid region and $0$ in the vapor region. Eq. (\ref{Eq:ghost_velocity}) is only employed in a narrow band region near the interface, which will be discussed in the next section. Note that the ghost velocities will be set for both the cell-centered and the face-centered velocities. After that, we solve Eqs. (\ref{Eq:discretization_1})-(\ref{Eq:discretization_3}) separately for $\mathbf{u}_{f,liq}^*$ and $\mathbf{u}_{f,vap}^*$. Then, for the projection step, Eq. (\ref{Eq:Poisson_equation}) is modified to
\begin{equation}
     \nabla_c \cdot \left(\frac{\Delta t}{\rho^{n+\frac{1}{2}}_f}\nabla p^{n+1} \right) = \left\{ 
     \begin{aligned}
        &\nabla_c \cdot \mathbf{u}_{f,liq}^* \quad \text{if $f_C = 1$}, \\
        &\nabla_c \cdot \mathbf{u}_{f,vap}^* \quad \text{if $f_C = 0$}. \\  
     \end{aligned}
     \right.
     \label{Eq:poisson_ghost_fluid}
\end{equation}
Note that there is no singular source term on the right-hand side of the mass conservation equation, compared with the one-fluid method, since the velocity jump has been appropriately taken into account by Eq. (\ref{Eq:ghost_velocity}) in the ghost fluid method. With $p^{n+1}$ solved, the face-centered velocity is finally updated by
\begin{equation}
    \begin{aligned}
      \mathbf{u}_{f,liq}^{n+1} &= \mathbf{u}_{f,liq}^* - \frac{\Delta t}{\rho^{n+\frac{1}{2}}}\nabla p^{n+1}&\quad \text{if $f_C = 1$}, \\
    \mathbf{u}_{f,vap}^{n+1} &= \mathbf{u}_{f,vap}^* - \frac{\Delta t}{\rho^{n+\frac{1}{2}}}\nabla p^{n+1}&\quad \text{if $f_C = 0$}. \\    
    \end{aligned}
\end{equation}
The cell-centered velocity is computed accordingly using Eq. (\ref{Eq:discretization_5}). Subsequently, we introduce how to obtain the interfacial velocity at the markers. As illustrated in Fig. \ref{Fig:intefacial_vel_calculation}, for a given marker $i$, its velocity is determined using the values at the cell centers of the nearest background grid cells 1-4. Initially, $\mathbf{u}_{\Gamma,1-4}$ are computed by applying the jump condition
\begin{equation}
     \mathbf{u}_\Gamma = \left\{ 
     \begin{aligned}
        &\mathbf{u}_{liq} - \frac{\dot{m}}{\rho_{liq}}\mathbf{n} \quad \text{if $f_C = 1$}, \\
        &\mathbf{u}_{vap} - \frac{\dot{m}}{\rho_{vap}}\mathbf{n} \quad \text{if $f_C = 0$}. \\  
     \end{aligned}
     \right.
     \label{Eq:jump_interfacial_vel}
\end{equation}
at the cell centers. Then, at the marker $i$, $\mathbf{u}_{\Gamma,i}$ is calculated via bilinear interpolation,
\begin{equation}
\mathbf{u}_{\Gamma,i} = (1 - \overline{x})(1 - \overline{y})\mathbf{u}_{\Gamma,2} + (1 - \overline{x})\overline{y}\mathbf{u}_{\Gamma,1}+ \overline{x}(1 - \overline{y})\mathbf{u}_{\Gamma,3}+\overline{x} \overline{y}\mathbf{u}_{\Gamma,4},
\end{equation}
where $\overline{x}$ and $\overline{y}$ are computed by
\begin{equation}
\overline{x} = \frac{x_i - x_2}{\Delta x}\quad\text{and}\quad\overline{y} = \frac{y_i - y_2}{\Delta y},
\end{equation}
respectively.
\subsection{Solving the energy equation}
\begin{figure}[htbp]
    \centering
    \includegraphics[width=1.0\textwidth]{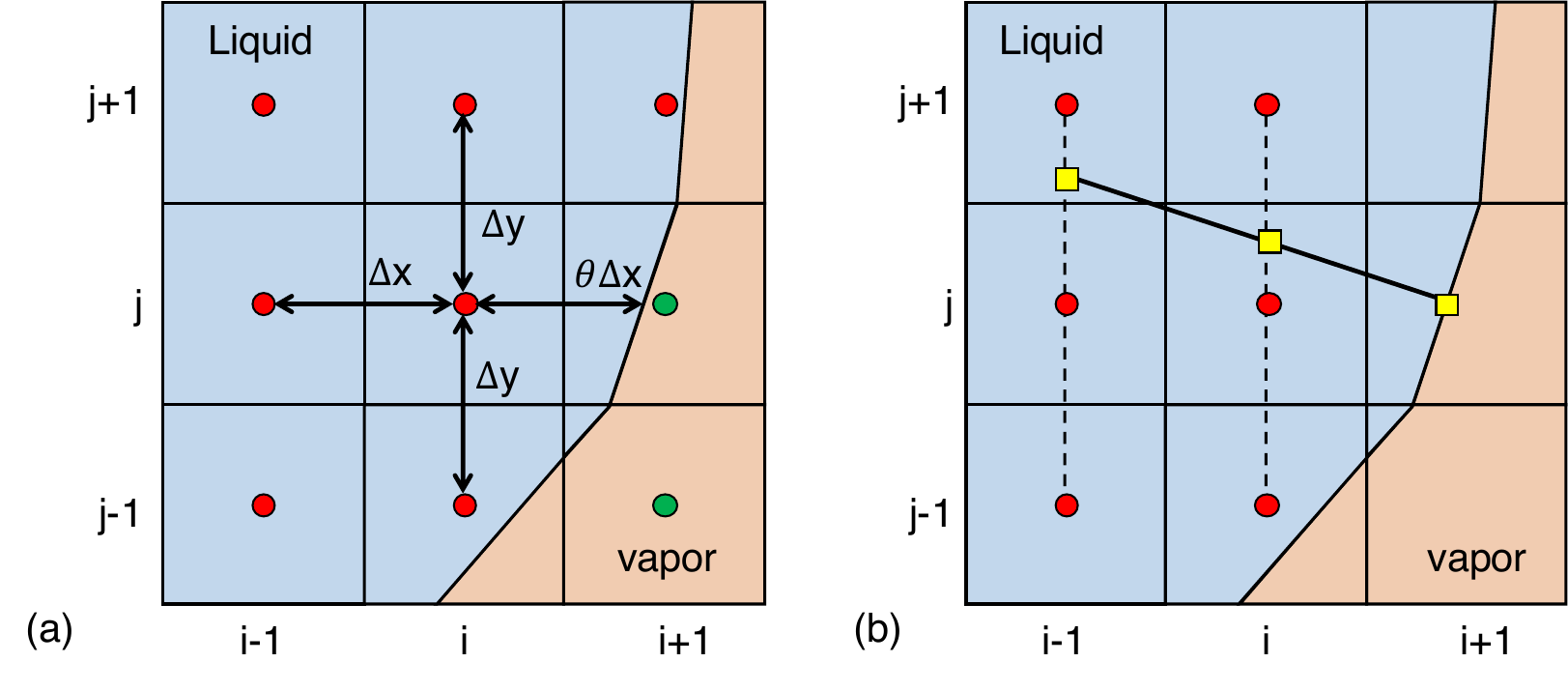}
    \caption{(a) The discretization stencil of $\nabla^2 T$ for a cell near the interface. (b) The computation of the normal temperature gradient at the liquid side for an interfacial cell.}
    \label{Fig:stencil}
\end{figure}
In this section we elaborate the solution of the energy equation. We use the finite difference method to compute the temperature field separately in the liquid domain and in the vapor domain:
\begin{align}
    &\rho_k C_{p,k} \left( \frac{T_k^* - T_k^{n-\frac{1}{2}}}{\Delta t} + \mathbf{u}_k^n \cdot \nabla T_k^{n-\frac{1}{2}}\right) = 0,\label{Eq:energy_advection} \\
    &\rho_k C_{p,k} \left( \frac{T_k^{n+\frac{1}{2}} - T_k^*}{\Delta t} \right)  = \nabla \cdot (\lambda \nabla T_k^{n+\frac{1}{2}}),\label{Eq:energy_diffusion}
\end{align}
where $k = liq$ or $k = vap$ denotes the current region of interest. The advection term is discretized by using a third-order accurate WENO scheme \cite{jiang1996efficient}, and the diffusion term is solved implicitly. When solving the energy equation, we need to correctly impose the Dirichlet boundary condition \cite{sato2013sharp}
\begin{equation}
    T_\Gamma = T_{sat},
    \label{Eq:temp_bc_interface}
\end{equation}
at the interface, where $T_{sat}$ is the saturation temperature. This is implemented in a similar manner as in Ref. \cite{gibou2007level}. The main principle is to fill the uncompleted discretization stencil with an extrapolated value by considering $T_\Gamma = T_{sat}$. For example, by using the standard second-order accurate central difference scheme for a cell $(i,j)$ which is far from the interface, $T_{xx}$ in $\nabla^2 T$ can be written as
\begin{equation}
    T_{xx} = \frac{\frac{T_{i+1} - T_{i}}{\Delta x} - \frac{T_{i} - T_{i-1}}{\Delta x}}{\Delta x}.
    \label{Eq:Txx_normal}
\end{equation}
When the cell is near the interface, sometimes the stencil cell may across the interface, as shown in Fig. \ref{Fig:stencil}(a). The ghost temperature $T^{ghost}$ for the cell $(i+1,j)$ in Fig. \ref{Fig:stencil}(a) can be extrapolated by
\begin{equation}
    T^{ghost}_{i+1} = T_\Gamma + \frac{T_\Gamma - T_{i}}{\theta}(1 - \theta),
    \label{Eq:ghost_T}
\end{equation}
where $\theta  = (x_\Gamma  - x_i) / \Delta x$, ranging from $0$ to $1$. Substituting Eq. (\ref{Eq:ghost_T}) into Eq. (\ref{Eq:Txx_normal}) gives 
\begin{equation}
    T_{xx} = \frac{\frac{T_{\Gamma} - T_{i}}{ \theta \Delta x} - \frac{T_{i} - T_{i-1}}{\Delta x}}{\Delta x},
    \label{Eq:Txx_ghost}
\end{equation}
leading to a second-order accurate symmetric discretization \cite{gibou2002second}.  

As shown in Eq. (\ref{Eq:energy_advection}) and Eq. (\ref{Eq:energy_diffusion}), the solution of the energy equation is split into two steps, utilizing an intermediate temperature field $T^*$ for the contribution of the advection terms. This strategy is employed because, during one time step, the cell centers may be swept by the interface, leading to unknown temperature values inside the newly created vapor/liquid cells \cite{tanguy2014benchmarks, gibou2007level}. By solving the advection equation Eq. (\ref{Eq:energy_advection}) with $T^*$ and extrapolating it to the ghost region, the unknown values are determined. The extrapolation \cite{tanguy2014benchmarks} is achieved by iteratively solving  
\begin{equation}
\begin{aligned}
    &\frac{\partial f}{\partial \tau} + (1 - f_C) \mathbf{n}\cdot\nabla f = 0 &\quad\text{from liquid to vapor,} \\
    &\frac{\partial f}{\partial \tau} - f_C\mathbf{n}\cdot\nabla f = 0 &\quad\text{from vapor to liquid,}
\end{aligned}
     \label{Eq:extrapolation}
\end{equation}
where $\tau$ is a pseudo time, and $f$ the variable to be extended. Instead of solving Eq. (\ref{Eq:extrapolation}) in the whole domain, we solve it within a narrow band near the interface. A cell is defined as being in the narrow band if it or any of its neighbors is cut by the interface.

After solving the energy equation, the mass flux $\dot{m}$ is calculated as
\begin{equation}
    \dot{m} = \frac{\left[q\right]}{h_{lg}} = \frac{\lambda_{liq} \nabla T_{liq} \cdot \mathbf{n} - \lambda_{vap} \nabla T_{vap} \cdot \mathbf{n}}{h_{lg}},
    \label{Eq:mass_flux_computation}
\end{equation}
where $h_{lg}$ is the latent heat, and $q_{liq/vap} = \lambda_{liq/vap} \nabla T_{liq/vap} \cdot \mathbf{n}$ represents the heat flux at each side of the interface. As the temperature gradient is not continuous across the interface \cite{zhao2022boiling}, to evaluate the heat flux accurately and robustly, we compute the normal temperature gradients in the two phases separately by adopting an Embedded Boundary Method \cite{johansen1998cartesian}. The computation in the liquid side is sketched in Fig. \ref{Fig:stencil}(b). Two points are found along the normal direction at first, where the temperature values are obtained from neighboring cells using bi-quadratic interpolation. Then the normal temperature gradient at the interface is interpolated by 
\begin{equation}
    \frac{\partial T}{\partial n} = \frac{1}{d_2 - d_1}(\frac{d_2}{d_1}(T_\Gamma - T_1) - \frac{d_1}{d_2}(T_\Gamma - T_2)),
    \label{Eq:normal_gradient}
\end{equation}
where $d_1$ and $d_2$ denote the distances from the interface centroid to the two points, respectively. It is noteworthy that the boundary condition at the interface, as specified in Eq. (\ref{Eq:temp_bc_interface}), is directly imposed in Eq. (\ref{Eq:normal_gradient}). Once the heat fluxes at the interface are determined, the mass flux $\dot{m}$ in the cells cut by the interface is obtained. To compute the ghost fluid velocities, the mass flux is then extended to the entire narrow band \cite{tanguy2014benchmarks} by solving Eq. (\ref{Eq:extrapolation}).   

\subsection{Overall numerical process}
For clarity, the numerical procedures for each time iteration are outlined as below:
\begin{enumerate}
\item Solve the advection part of the energy equation Eq. (\ref{Eq:energy_advection}).
\item Advect the interface using the EBIT method, with the interfacial velocity obtained from Eq. (\ref{Eq:jump_interfacial_vel}).
\item Solve the diffusion part of the energy equation Eq. (\ref{Eq:energy_diffusion}).
\item Compute the mass flux and accordingly set the ghost velocities using Eq. (\ref{Eq:ghost_velocity}).
\item Solve the mass and momentum equations Eqs. (\ref{Eq:discretization_1})-(\ref{Eq:discretization_5}).
\end{enumerate}

\section{Numerical Results}
\label{sec3}
In this section, the proposed method is validated through a series of numerical benchmark tests. The codes developed for this work, as well as the configurations for all simulations, are freely available in the Basilisk sandbox \cite{tiansandbox}. A quad/octree based adaptive mesh reﬁnement (AMR) technique \cite{popinet2009accurate} is employed to improve computational efficiency. For a given grid level $L$, the corresponding number of cells in one direction for a cubic domain is $N = 2^{L}$. In the following tests, AMR is utilized, except for the one-dimensional problems. The grid levels reported in each case represent the highest refinement level and correspond to the smallest grid cell size used. In cases involving a non-cubic domain, it is important to note that $2^{L}$ always corresponds to the number of cells employed in the direction with the maximum length. 

\subsection{Stefan problem}
\label{sec3.1}
\begin{figure}[htbp]
    \centering
    \includegraphics[width=1.0\textwidth]{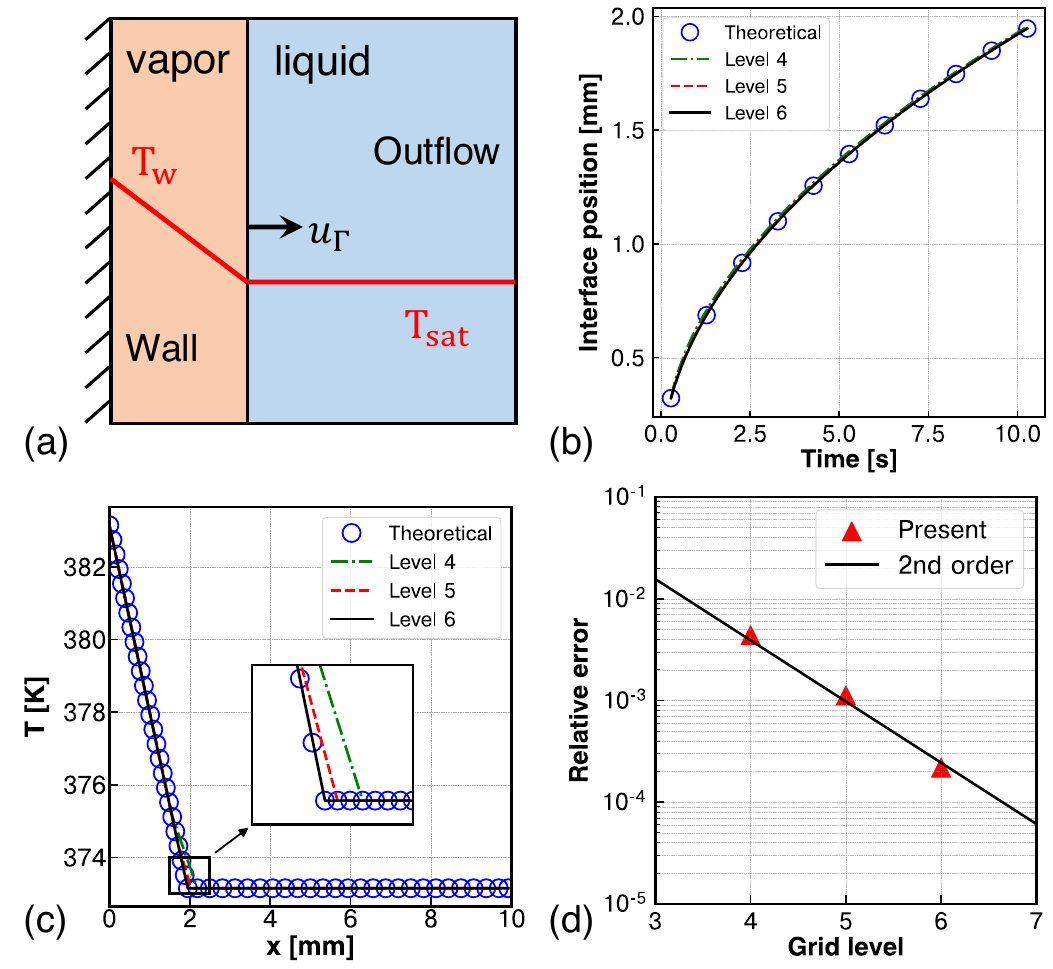}
    \caption{Stefan problem: (a) Schematic of the 1D Stefan problem. (b) Time history of the interface position. (c) Temperature distribution at \( t = 10.282 \) s. (d) Relative error of the interface position on different grid resolutions. Grid levels 4 to 6 correspond to effective grid resolutions ranging from \( 16 \times 1 \) cells to \( 64 \times 1 \) cells, resulting in minimum grid sizes from $625\ \rm{\mu m}$ to $156.25\ \rm{\mu m}$.}
    \label{Fig:stefan_problem}
\end{figure}
\begin{figure}[htbp]
    \centering
    \includegraphics[width=1.0\textwidth]{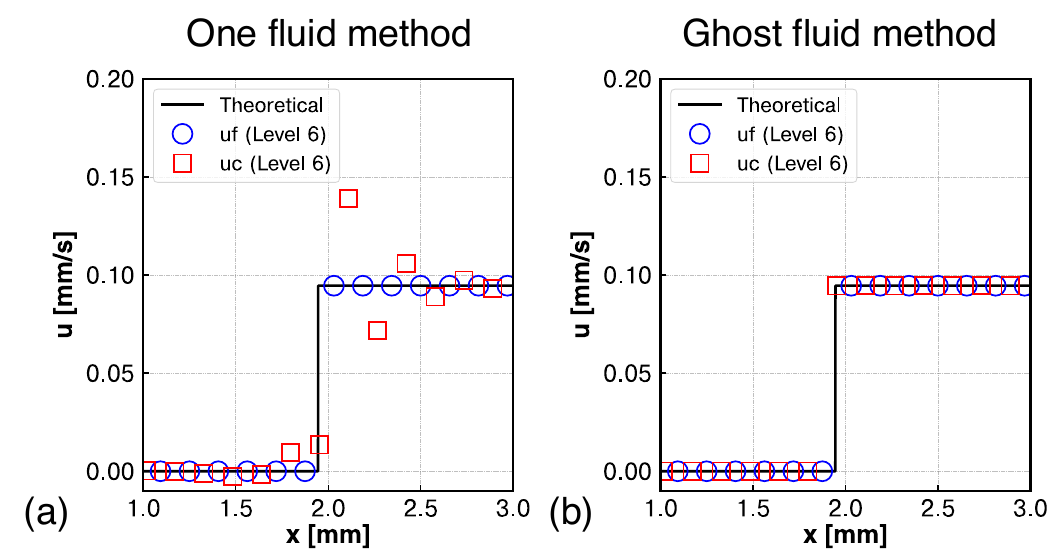}
    \caption{Stefan problem: The cell-centered velocity uc and face-centered velocity uf distributions across the interface at \( t = 10.282 \) s, obtained by the one fluid method (a) and the ghost fluid method (b).}
    \label{Fig:stefan_problem_vel}
\end{figure}
\begin{figure}[htbp]
    \centering
    \includegraphics[width=0.8\textwidth]{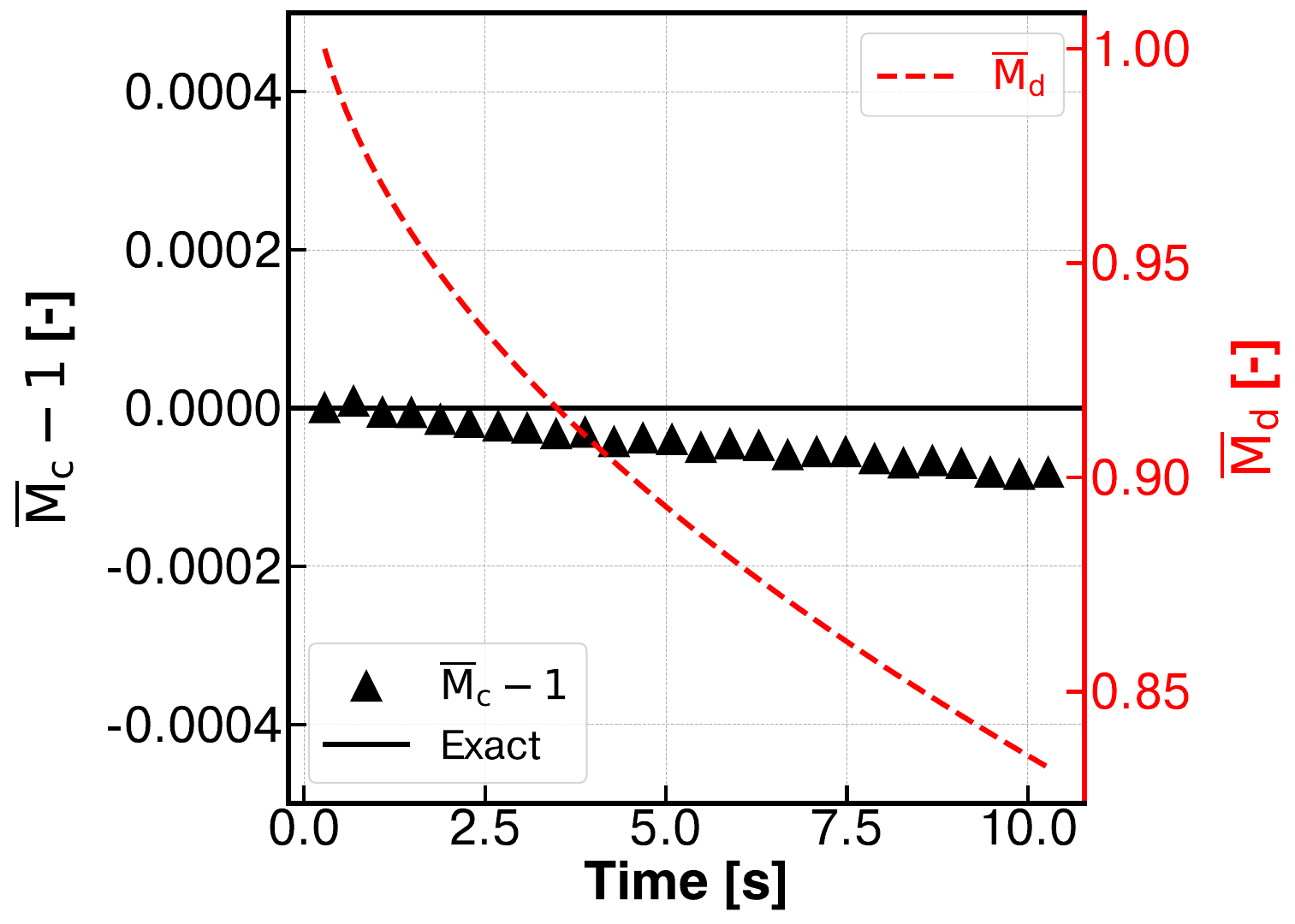}
    \caption{Relative error in mass conservation at the highest grid level for the Stefan problem: $\overline{M}_d(t)$ represents the mass within the computational domain, and $\overline{M}_c(t)$ represents the mass considering outflow. Both are non-dimensionalized by the initial mass in the computational domain, $M_d(0)$.}
    \label{Fig:stefan_conservation}
\end{figure}
Here, the 1D Stefan problem is considered, which is widely used to verify phase change models \cite{malan2021geometric, sato2013sharp}. As shown in Fig. \ref{Fig:stefan_problem}(a), a thin vapor layer is placed between a liquid at saturation temperature $T_{sat}$ and a heated wall with temperature $T_w$. The left boundary (at $x = 0$) is a solid wall, while the right boundary (at $x = l$) is set as an outlet. In this case, the vapor layer grows over time due to boiling caused by the heat flux from the vapor side, pushing the liquid out of the domain. An analytical solution is available \cite{sato2013sharp}, and the interface position $X_\Gamma(t)$ is given by 
\begin{equation}
    X_\Gamma(t) = 2 \beta \sqrt{\alpha_v t},
    \label{Eq:stefan_interface_position}
\end{equation}
where $\alpha_v = \lambda_{vap} / \rho_{vap} C_{p,vap}$ represents the thermal diffusivity of the vapor, and $\beta$ is the solution of a transcendental equation:
\begin{equation}
    \beta \mathrm{exp}(\beta^2) \mathrm{erf}(\beta)  = \frac{C_{p,vap}(T_w - T_{sat})}{\sqrt{\pi} h_{lg}}.
\end{equation}
During the simulation, the liquid remains at the saturation temperature, while the theoretical temperature distribution within the vapor layer is given by
\begin{equation}
    T(x,t) = T_w + \frac{T_{sat} - T_w}{\mathrm{erf}(\beta)}\mathrm{erf}(\frac{x}{2\sqrt{\alpha_v t}}).
\end{equation}
This problem can be characterized by the Jacob number $\mathrm{Ja}$, given by
\begin{equation}
    \mathrm{Ja} = \frac{\rho_{liq}C_{p,liq} \Delta T}{\rho_{vap} h_{lg}}, 
\end{equation}
where $\Delta T = T_w - T_{sat}$ is the superheat.

Following Ref. \cite{malan2021geometric}, the domain length $l$ for this problem is set to $10\ \rm{mm}$. With the Jakob number being $\mathrm{Ja} = 29.84$, the physical properties are set as
\begin{equation}
\left\{\begin{aligned}
\rho_{liq} & =958\ \mathrm{kg/m^3}, \mu_{liq}=2.82\times10^{-4}\ \mathrm{Pa\cdot s},\\ \lambda_{liq}&=0.68\ \mathrm{W/m K}, C_{p,liq} =4216\ \mathrm{J/kg\cdot K}, \\
\rho_{vap} & =0.6\ \mathrm{kg/m^3}, \mu_{vap}=1.23\times10^{-5}\ \mathrm{Pa\cdot s}, \\
\lambda_{vap} &= 0.025\ \mathrm{W/m\cdot K}, C_{p,vap} =2080\ \mathrm{J/kg\cdot K}, \\
T_{sat} & =373.15\ \mathrm{K}, T_w = 383.15\ \mathrm{K}, h_{lg} = 2.256 \times 10^6\ \mathrm{J/kg}.
\end{aligned}\right.
\end{equation}
 Initially, the thickness of the vapor layer is $322.5\ \rm{\mu m}$, corresponding to the theoretical solution Eq. (\ref{Eq:stefan_interface_position}) at $t = 0.282\ \rm{s}$. The simulation is performed up to $t = 10.282\ \rm{s}$, and three grid levels, ranging from 4 to 6, are used. In Figs. \ref{Fig:stefan_problem}(b) and \ref{Fig:stefan_problem}(c), the interface positions and the final temperature distributions, obtained with different grid levels, are compared with the theoretical solutions. Excellent agreement is observed for all grid resolutions. The relative error of the final interface position can be calculated by
\begin{equation}
E(X_{\Gamma}) = \frac{|X_{\Gamma,theo} - X_{\Gamma,num}|}{X_{\Gamma,theo}},
\end{equation}
where the subscripts $theo$ and $num$ represent the theoretical solution and the numerical result, respectively. Fig. \ref{Fig:stefan_problem}(d) shows the relative errors for different grid levels, indicating a second-order convergence rate. Moreover, to measure the accuracy of mass conservation, we follow the method of Sato and Ni\v{c}eno \cite{sato2013sharp} by computing two types of mass: the mass within the computational domain,
\begin{equation}
M_d (t)=\sum \rho V,
\label{Eq:md}
\end{equation}
and the mass considering the outflow
\begin{equation}
M_c(t)=\sum \rho V+\sum_{0}^{t} \sum_{\text {Outlet faces }} \rho \mathbf{u} \cdot \mathbf{S} \Delta t,
\label{Eq:mc}
\end{equation}
where $\mathbf{S}$ represents the area vector of the cell faces belonging to the outlet boundary. The time histories of the two dimensionless masses, $\overline{M}_d$ and $\overline{M}_c$, non-dimensionalized by the initial mass within the computational domain, $M_d(0)$, are given in Fig. \ref{Fig:stefan_conservation} for the simulation at the finest grid level. It is shown that $\overline{M}_d$ decreases due to the volume expansion of vapor, while the maximum relative conservation error, $\overline{M}_c - 1$, is $0.0082 \%$. The reasons for the conservation error being larger than machine zero, which is also observed in the following 2D simulations, are twofold: Firstly, the EBIT method is a Front-Tracking method in which exact mass conservation is not assured \cite{tryggvason2011direct}. Secondly, in the 2D tests, the volume fraction is computed based on a piecewise linear reconstruction of interface segments \cite{pan2023edge}, which introduces errors in the calculation of density. For Front-Tracking methods, the accuracy of mass conservation can be improved by using high-order time-integration methods for marker advection \cite{tryggvason2011direct} or by employing specialized methods to interpolate marker velocities \cite{gorges2022reducing}, which is, however, beyond the scope of the current work.

Additionally, we also examine the velocity distribution to further validate our method. The velocity distributions obtained by the one fluid method and the ghost fluid method are plotted in Fig. \ref{Fig:stefan_problem_vel}. It is shown that a sharp velocity jump appears at the interface due to phase change. With the one fluid method, unphysical oscillations are observed for $\mathbf{u}_c$ as it is approximately projected \cite{zhao2022boiling}. In contrast, by using the ghost fluid method, the sharp velocity jump is accurately captured without introducing oscillations near the interface. As discussed earlier, the discontinuity at the interface is removed by setting the ghost velocity according to the jump condition Eq. (\ref{Eq:ghost_velocity}), thus improving numerical stability.

\subsection{Sucking problem}
\label{sec3.2}
\begin{figure}[htbp]
    \centering
    \includegraphics[width=1.0\textwidth]{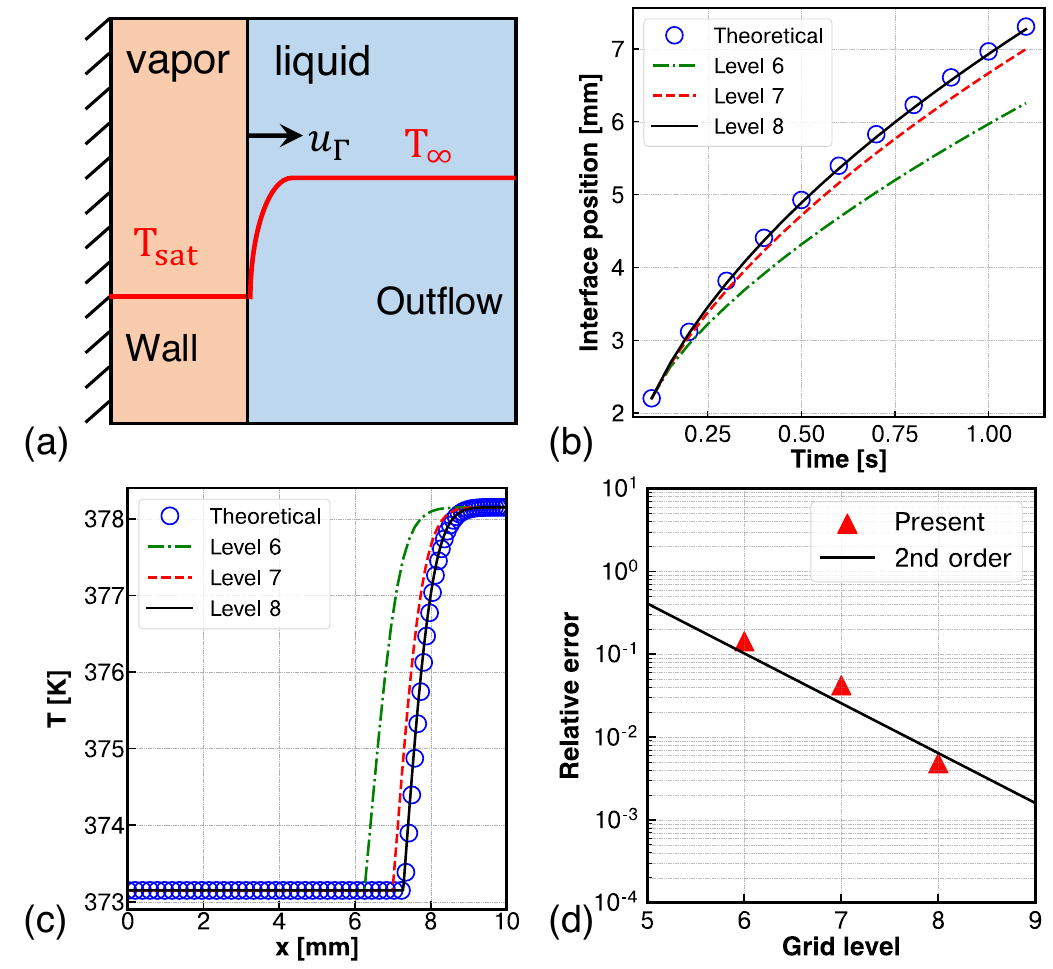}
    \caption{Sucking problem: (a) Schematic of the 1D sucking problem. (b) Time history of the interface position. (c) Temperature distribution at \( t = 1.1 \) s. (d) Relative error of the interface position on different grid resolutions.
  Grid levels 6 to 8 correspond to effective grid resolutions ranging from \( 64 \times 1 \) cells to \( 256 \times 1 \) cells, resulting in minimum grid sizes from $156.25\ \rm{\mu m}$ to $39.06\ \rm{\mu m}$.}
    \label{Fig:sucking_problem}
\end{figure}
\begin{figure}[htbp]
    \centering
    \includegraphics[width=0.8\textwidth]{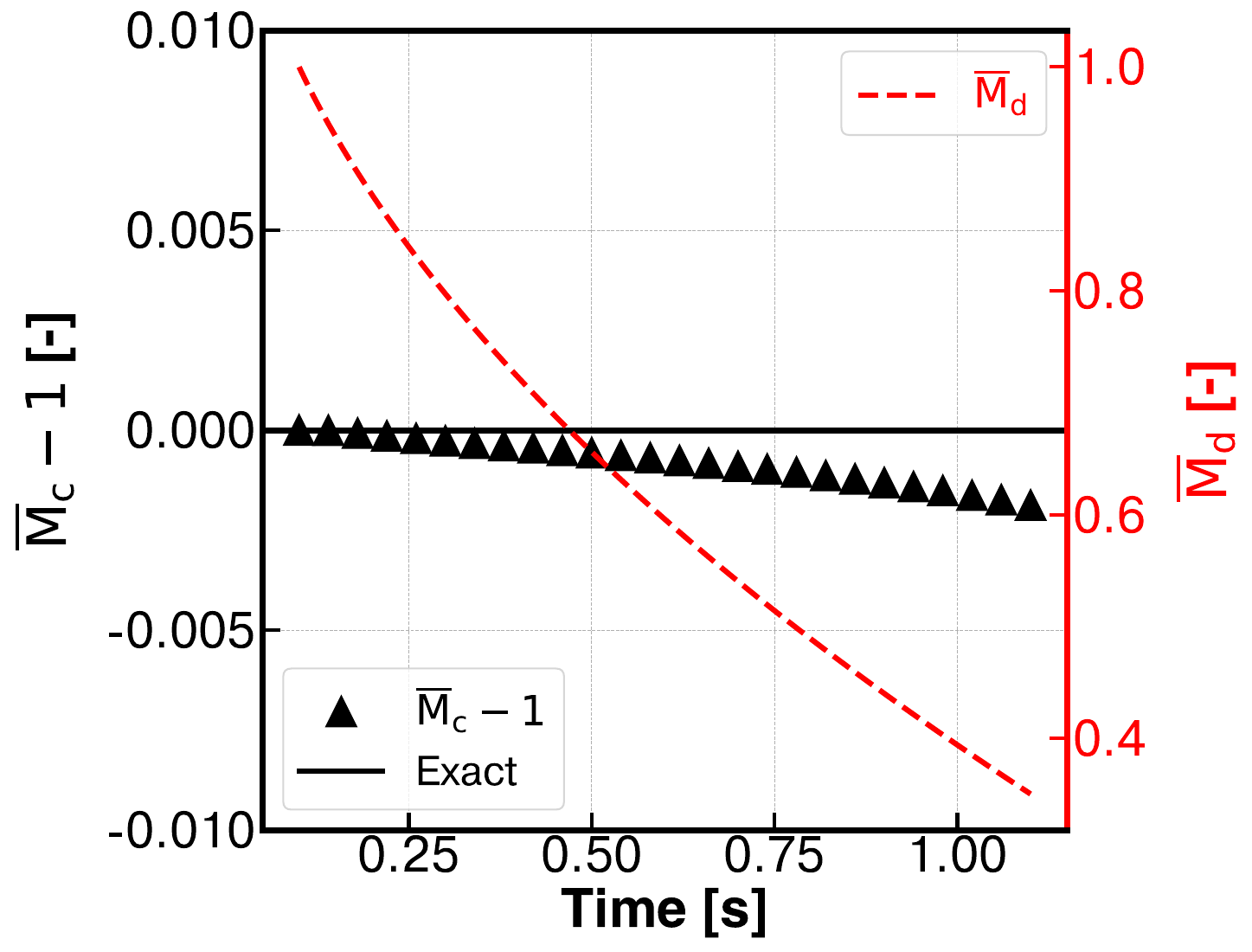}
    \caption{Relative error in mass conservation at the highest grid level for the sucking problem.}
    \label{Fig:sucking_conservation}
\end{figure}
Proposed by Welch and Wilson \cite{welch2000volume}, the 1D sucking problem is another popular benchmark test used to validate phase change models under superheated conditions. As shown in Fig. \ref{Fig:sucking_problem}(a), a superheated liquid with temperature $T_{\infty}$ is adjacent to a vapor layer at saturation temperature $T_{sat}$. The left boundary (at $x = 0$) is a solid wall with the temperature fixed at $T_{sat}$, and the right boundary (at $x = l$) is an outlet. Due to the superheat, the liquid starts to boil, causing the volume of vapor to expand and push the interface rightwards. Compared with the Stefan problem, the sucking problem is more challenging because of the formation of a thin thermal boundary layer close to the interface on the liquid side. For this problem, a theoretical solution can be obtained from the similarity solution \cite{boyd2023consistent}. The analytical solution for the interface position is given by 
\begin{equation}
    X_\Gamma(t) = 2 \beta \sqrt{\alpha_l t},
    \label{Eq:sucking_interface_position}
\end{equation}
where $\alpha_l = \lambda_{liq} / \rho_{liq} C_{p,liq}$ is the thermal diffusivity of the liquid, and the growth constant $\beta$ is obtained by solving the equation
\begin{equation}
    e^{\beta^2}\mathrm{erf}(\beta)\left[ \beta - \frac{(T_\infty - T_{sat})C_{p,vap}\lambda_{liq}\sqrt{\alpha_v}e^{-\left(\beta\frac{\rho_{vap}\sqrt{\alpha_v}}{\rho_{liq}\sqrt{\alpha_l}}\right)^2}}{h_{lg} \lambda_{vap} \sqrt{\pi \alpha_l}\mathrm{erfc}\left(\beta\frac{\rho_{vap}\sqrt{\alpha_v}}{\rho_{liq}\sqrt{\alpha_l}}\right)}\right] = 0.
    \label{Eq:sucking_gamma}
\end{equation}
The theoretical solution for the liquid temperature distribution is accordingly given by
\begin{equation}
    T(x,t) = T_\infty - \frac{(T_\infty - T_{sat})}{\mathrm{erfc}\left(\beta\frac{\rho_{vap}\sqrt{\alpha_v}}{\rho_{liq}\sqrt{\alpha_l}}\right)}  \mathrm{erfc}\left(\frac{x}{2\sqrt{\alpha_l t}} + \frac{\beta (\rho_{vap} - \rho_{liq})\sqrt{\alpha_v}}{\rho_{liq}\sqrt{\alpha_l}}\right).
\end{equation}
Following Ref. \cite{zhao2022boiling}, the physical properties for this problem are set as
\begin{equation}
\left\{\begin{aligned}
\rho_{liq} & =958.4\ \mathrm{kg/m^3}, \mu_{liq}=2.80\times10^{-4}\ \mathrm{Pa\cdot s},\\ \lambda_{liq}&=0.679\ \mathrm{W/m K}, C_{p,liq} =4216\ \mathrm{J/kg\cdot K}, \\
\rho_{vap} & =0.597\ \mathrm{kg/m^3}, \mu_{vap}=1.26\times10^{-5}\ \mathrm{Pa\cdot s}, \\
\lambda_{vap} &= 0.025\ \mathrm{W/m\cdot K}, C_{p,vap} =2030\ \mathrm{J/kg\cdot K}, \\
T_{sat} & =373.15\ \mathrm{K}, T_{\infty} = 383.15\ \mathrm{K}, h_{lg} = 2.26 \times 10^6\ \mathrm{J/kg}.
\end{aligned}\right.
\end{equation}
With the superheat $\Delta T = T_\infty - T_{sat} =  10\ \mathrm{K}$, the Jakob number is $\mathrm{Ja} = 29.95$.

In this case, the domain length $l$ is $10\ \rm{mm}$. Initially the interface is located at $x = 0.476\ \rm{mm}$, which corresponds to the theoretical solution at $t = 0.1\ \rm{s}$. We initialize the temperature field according to the theoretical solution at $t = 0.1\ \rm{s}$ and conduct the simulation up to $t = 1.1\ \rm{s}$. The interface positions, obtained with different grid levels ranging from 6 to 8, are plotted against the simulation time in Fig. \ref{Fig:sucking_problem}(b). It is shown that the results converge to the analytical solution with increasing grid refinement. The relative errors of the final interface position at $t = 1.1\ \rm{s}$ are shown in Fig. \ref{Fig:sucking_problem}(d), from which a second-order convergence rate can be observed. Furthermore, in Fig. \ref{Fig:sucking_problem}(c), the temperature distributions at the end of the simulation are compared against the theoretical solution. The thermal boundary layer is well resolved for the simulation at grid level 8. Additionally, the time evolutions of $\overline{M}_d$ and $\overline{M}_c$, as defined in Eqs. (\ref{Eq:md}) and (\ref{Eq:mc}), are presented in Fig. \ref{Fig:sucking_conservation} for the simulation at grid level 8. As in the Stefan problem, $\overline{M}_d$ decreases due to liquid outflow, while $\overline{M}_c$ remains approximately 1. The maximum relative conservation error for the sucking problem is $0.19\%$.

\subsection{Bubble evaporation/condensation with a constant mass flux}
\label{sec3.3}
\begin{figure}[htbp]
    \centering
    \includegraphics[width=1.0\textwidth]{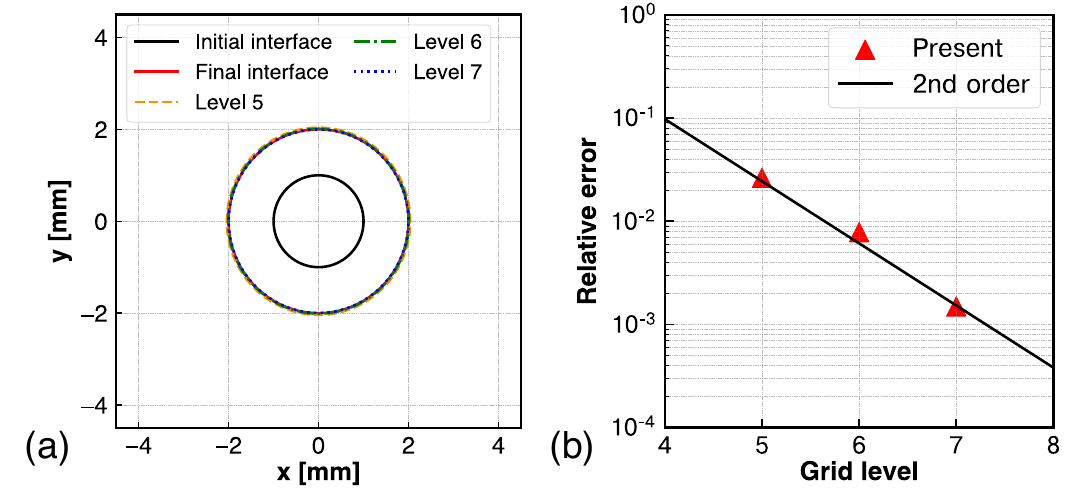}
    \caption{Bubble evaporation: (a) Comparison of the theoretical interface shape with the numerical results obtained with different grid levels. (b) Relative error of the bubble radius on different grid resolutions. Grid levels 5 to 7 correspond to effective grid resolutions ranging from \( 32 \times 32 \) cells to \( 128 \times 128 \) cells, resulting in minimum grid sizes from $250\ \rm{\mu m}$ to $62.5\ \rm{\mu m}$.}
    \label{Fig:fixed_evap}
\end{figure}
\begin{figure}[htbp]
    \centering
    \includegraphics[width=1.0\textwidth]{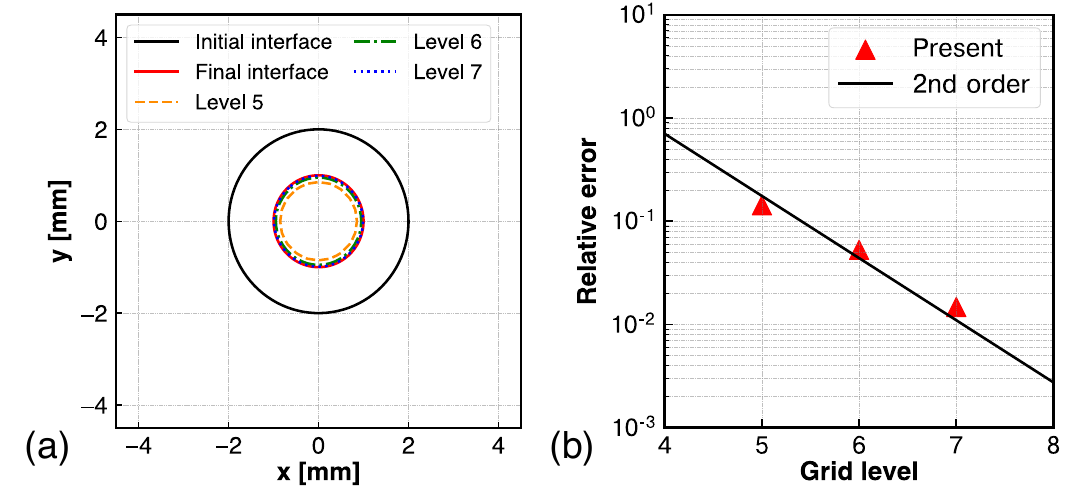}
    \caption{Bubble condensation: (a) Comparison of the theoretical interface shape with the numerical results obtained with different grid levels. (b) Relative error of the bubble radius on different grid resolutions. Grid levels 5 to 7 correspond to effective grid resolutions ranging from \( 32 \times 32 \) cells to \( 128 \times 128 \) cells, resulting in minimum grid sizes from $250\ \rm{\mu m}$ to $62.5\ \rm{\mu m}$.}
    \label{Fig:fixed_cond}
\end{figure}
\begin{figure}[htbp]
    \centering
    \includegraphics[width=1.0\textwidth]{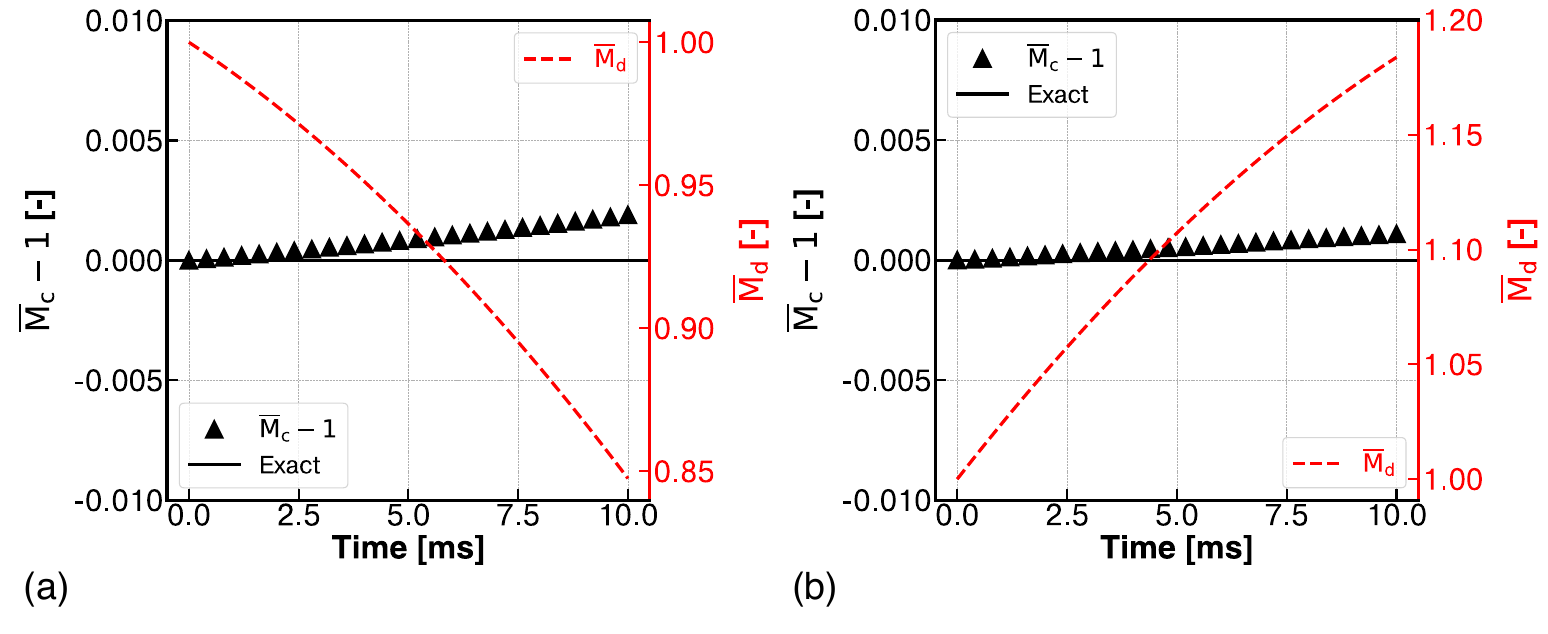}
    \caption{Relative error in mass conservation at the highest grid level for the bubble evaporation (a) and the bubble condensation (b).}
    \label{Fig:fixed_conservation}
\end{figure}
 A 2D bubble growing with a constant mass flux is considered in Ref. \cite{tanguy2014benchmarks} to verify the solution of the mass and momentum equations in the presence of phase change. Here, we also simulate this problem, additionally including a condensation setup. The computational domain is a square of $\left[ -4\ \rm{mm}, 4 \ \rm{mm} \right] \times \left[ -4\ \rm{mm}, 4 \ \rm{mm} \right]$, and the outflow boundary condition is employed for all the boundaries. The physical properties are given as
\begin{equation}
\left\{\begin{aligned}
&\rho_{liq}  =1000\ \mathrm{kg/m^3}, \mu_{liq}=1\times10^{-3}\ \mathrm{Pa\cdot s},\\
&\rho_{vap}  =1\ \mathrm{kg/m^3}, \mu_{vap}=1.26\times10^{-5}\ \mathrm{Pa\cdot s}, \\
&\sigma = 0.059\ \mathrm{N/m}.
\end{aligned}\right.
\end{equation}
Note that there is no need to solve the energy conservation equation, since the mass flux is manually imposed. We use $\dot{m} = 0.1\ \mathrm{kg/m^2 \cdot s}$ for the evaporation case and $\dot{m} = -0.1\ \mathrm{kg/m^2 \cdot s}$ for the condensation case. Theoretically, the bubble radius will evolve linearly with time \cite{tanguy2014benchmarks}:
\begin{equation}
    R(t) = R_0 + \frac{\dot{m}}{\rho_{vap}} t,
\end{equation}
where $R_0$ is the initial radius of the bubble, set to $0.1\ \rm{m}$ and $0.2\ \rm{m}$ for the evaporation and condensation cases, respectively. The simulations are performed with increasing grid levels from 5 to 7, and the final time is $t = 0.01\ \rm{s}$. Figs. \ref{Fig:fixed_evap}(a) and \ref{Fig:fixed_cond}(a) present the interfaces at the end of the simulation for different grid levels, which converge well to the theoretical solution. The circular shape of the bubble is well preserved for all grid levels. Additionally, the relative errors of the final bubble radius at different grid resolutions are plotted in Figs. \ref{Fig:fixed_evap}(b) and \ref{Fig:fixed_cond}(b), confirming the second-order convergence rate of the present method. In Fig. \ref{Fig:fixed_conservation}, the mass conservation at level 8 is examined by plotting the time histories of $\overline{M}_d$ and $\overline{M}_c$, as defined in Eqs. (\ref{Eq:mc}) and (\ref{Eq:md}). It is observed that $\overline{M}_d$ decreases in the evaporation case and increases in the condensation case due to the liquid outflow and inflow, respectively. The maximum relative conservation error is $0.19\%$ for the evaporation case and $0.11\%$ for the condensation case.

\subsection{Scriven problem}
\label{sec3.4}
\begin{figure}[htbp]
    \centering
    \includegraphics[width=1.0\textwidth]{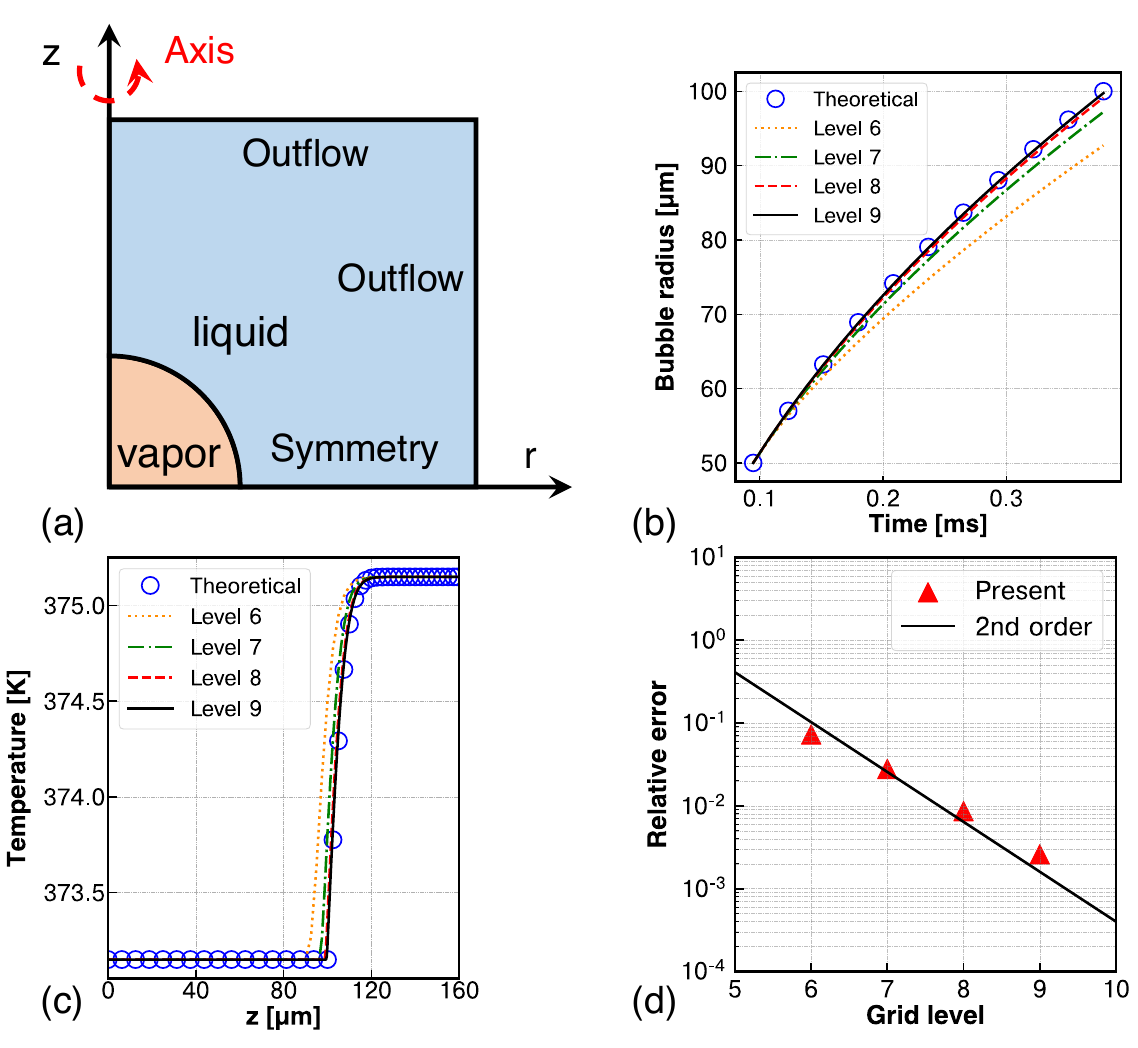}
    \caption{Scriven problem: (a) Schematic of the Scriven problem. (b) Time history of the bubble radius. (c) Temperature distribution along the z-axis at \( t = 10.282 \) s. (d) Relative error of the bubble radius on different grid resolutions. Grid levels 6 to 9 correspond to effective grid resolutions ranging from \( 64 \times 64 \) cells to \( 512 \times 512 \) cells, resulting in minimum grid sizes from $2.5\ \rm{\mu m}$ to $0.31\ \rm{\mu m}$.}
    \label{Fig:scriven}
\end{figure}
\begin{figure}[htbp]
    \centering
    \includegraphics[width=0.7\textwidth]{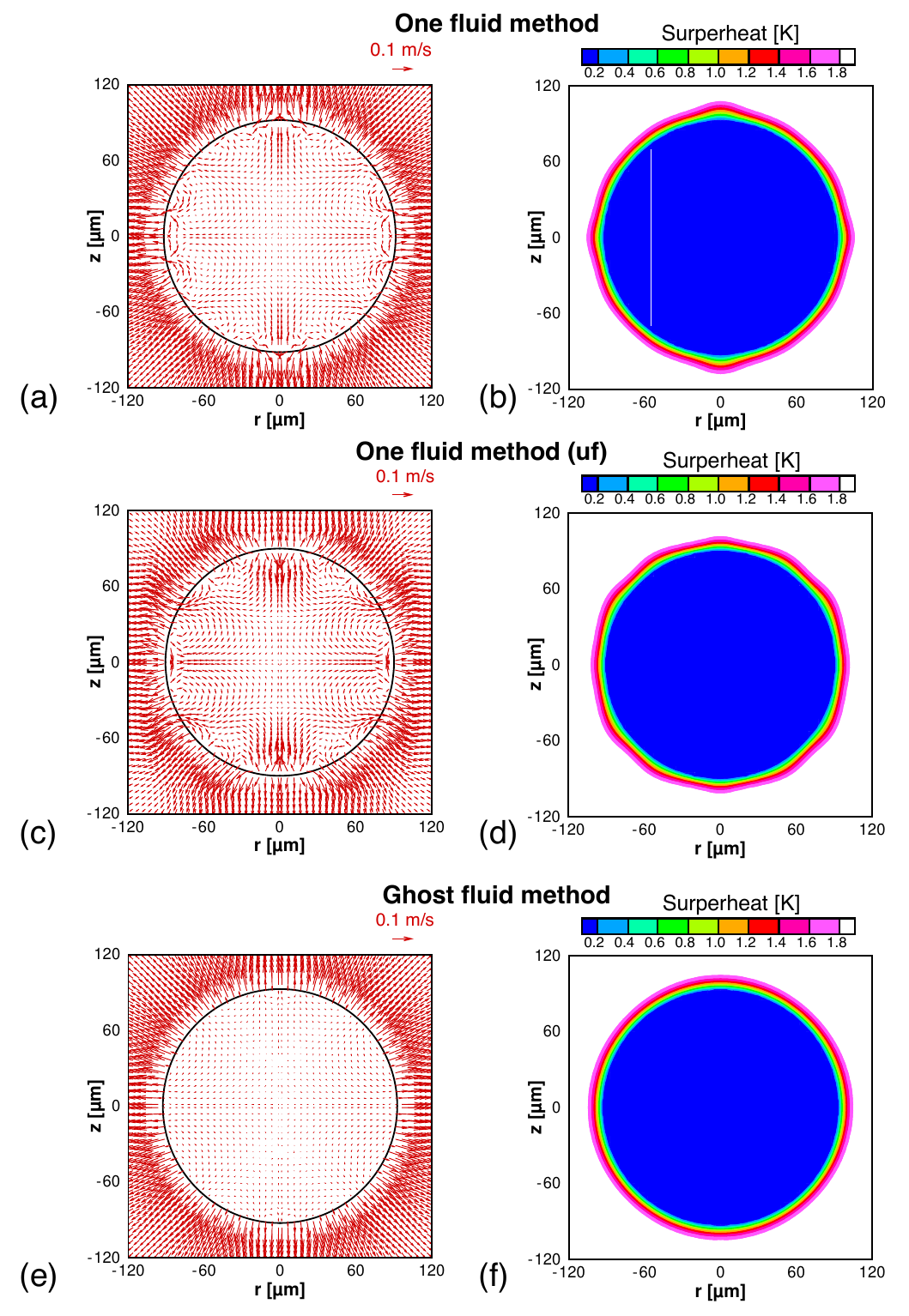}
    \caption{Scriven problem: The velocity vectors and superheat distributions ($T - T_{sat}$) obtained with different methods at the end of the simulation. (a)-(d) show the results with the one-fluid method. Note that (c)-(d) are obtained by adjusting the cell-centered velocity $\mathbf{u}_c$ to the average of the face-centered velocity $\mathbf{u}_f$ at each time step. (e)-(f) present the results with the ghost fluid method. The results are obtained at grid level 6. Only a quarter of the bubble is simulated in an axisymmetric configuration, and the results are mirrored about the axes $r = 0$ and $z = 0$ for better visualization.}
    \label{Fig:scrive_vel_superheat}
\end{figure}
\begin{figure}[htbp]
    \centering
    \includegraphics[width=0.8\textwidth]{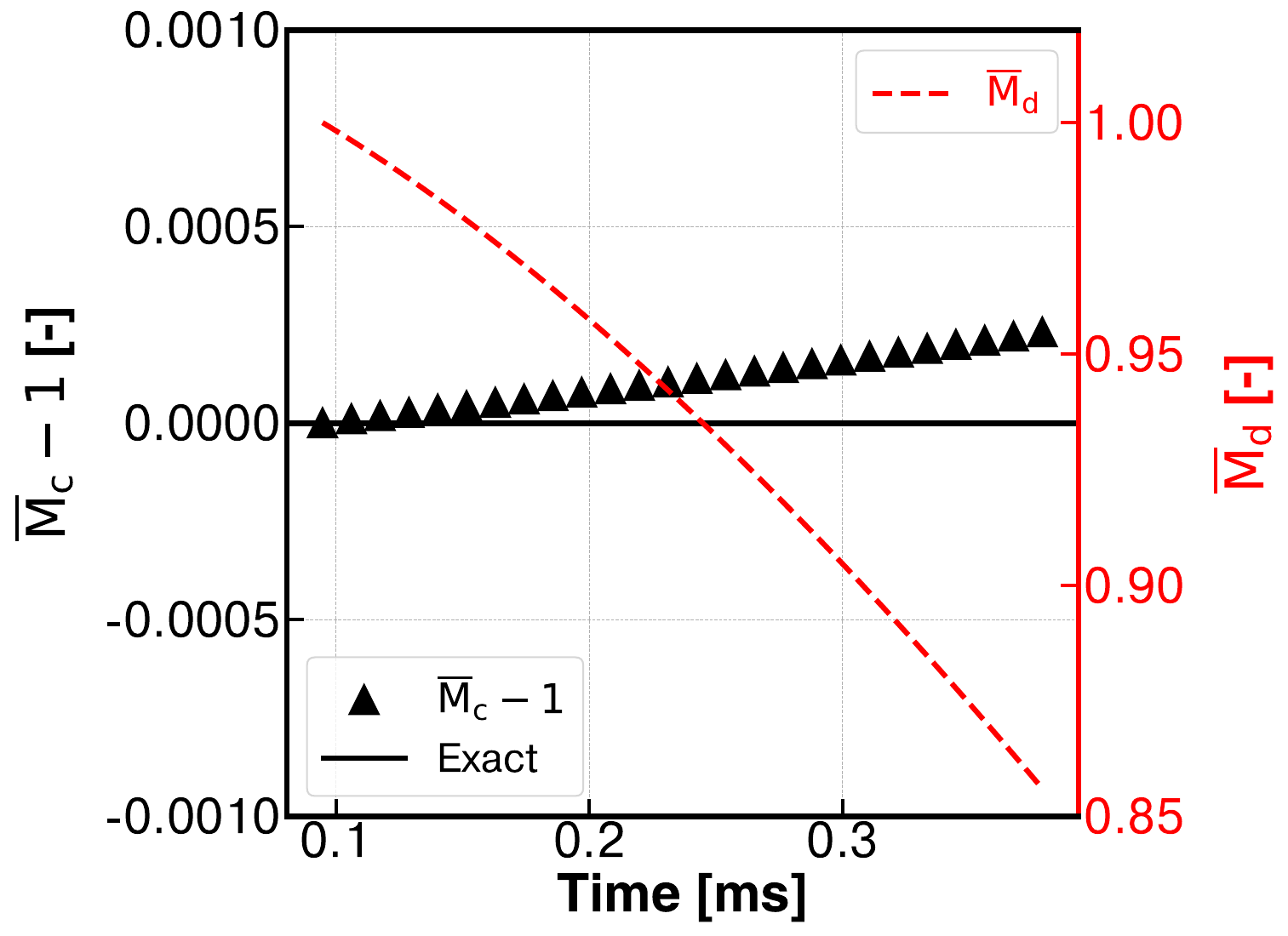}
    \caption{Relative error in mass conservation at the highest grid level for the Scriven problem.}
    \label{Fig:scriven_conservation}
\end{figure}
The growth of a static bubble at saturation temperature $T_{sat}$ in a superheated liquid with temperature $T_{\infty}$ is considered to validate the present method. This setup is known as the Scriven problem \cite{tanguy2014benchmarks,zhao2022boiling}, named after the theoretical solution obtained by Scriven \cite{scriven1995dynamics}. As illustrated in Fig. \ref{Fig:scriven}, it is computed in an axisymmetric configuration. The computational domain is a square with length $160\ \rm{\mu m}$. The left boundary is set as the axis of symmetry, and an outflow boundary condition is applied to the top and right boundaries. To improve computational efficiency, a symmetry boundary condition is employed on the bottom boundary, allowing the computation of only a quarter of the bubble. With the Jakob number $\mathrm{Ja} = 5.99$, the physical properties are set as
\begin{equation}
\left\{\begin{aligned}
\rho_{liq} & =958.4\ \mathrm{kg/m^3}, \mu_{liq}=2.80\times10^{-4}\ \mathrm{Pa\cdot s},\\ \lambda_{liq}&=0.679\ \mathrm{W/m K}, C_{p,liq} =4216\ \mathrm{J/kg\cdot K}, \\
\rho_{vap} & =0.597\ \mathrm{kg/m^3}, \mu_{vap}=1.26\times10^{-5}\ \mathrm{Pa\cdot s}, \\
\lambda_{vap} &= 0.025\ \mathrm{W/m\cdot K}, C_{p,vap} =2030\ \mathrm{J/kg\cdot K}, \\
T_{sat} & =373.15\ \mathrm{K}, T_{\infty} = 375.15\ \mathrm{K}, h_{lg} = 2.26 \times 10^6\ \mathrm{J/kg}, \sigma = 0.059 \mathrm{N/m}.
\end{aligned}\right.
\end{equation}

For this problem, the theoretical evolution of the bubble radius \cite{scriven1995dynamics} is given by
\begin{equation}
    R(t) = 2\beta\sqrt{\alpha_l t},
\end{equation}
where $\alpha_l = \lambda_{liq} / \rho_{liq} C_{p,liq}$ is the thermal diffusivity of the liquid, and the growth constant $\beta$ is obtained from solving the equation
\begin{equation}
\begin{aligned}
&\frac{\rho_{liq} C_{p,liq}\left(T_{\infty}-T_{sat}\right)}{\rho_{vap}\left(h_{lg}+\left(C_{p, liq}-C_{p,vap}\right)\left(T_{\infty}-T_{sat}\right)\right)}=\\
&2 \beta^2 \int_0^1 \exp \left(-\beta^2\left((1-\xi)^{-2}-2 (1 - \frac{\rho_{liq}}{\rho_{vap}}) \xi-1\right)\right) d \xi.
\end{aligned}
\label{Eq:scriven_growth_constant}
\end{equation}
During the bubble growth, the temperature inside the bubble remains at the saturation temperature $T_{sat}$ while the temperature distribution within the liquid can be theoretically calculated by
\begin{equation}
\begin{aligned}
T = & T_{\infty}-2 \beta^2\left(\frac{\rho_{vap}\left(h_{lg}+\left(C_{p, liq}-C_{p, vap}\right)\left(T_{\infty}-T_{sat}\right)\right)}{\rho_{liq} C_{p, liq}}\right) \\
& \int_{1-R / r}^1 \exp \left(-\beta^2\left((1-\xi)^{-2}-2\left(1-\frac{\rho_{vap}}{\rho_{liq}}\right) \xi-1\right)\right) d \xi.
\end{aligned}
\label{Eq:scriven_temperature}
\end{equation}
In the simulations, the bubble radius is initially set to $50\ \rm{\mu m}$, corresponding to the theoretical value at $t = 94.7\ \rm{\mu s}$. The temperature is accordingly initialized based on the theoretical solution Eq. (\ref{Eq:scriven_temperature}). This problem is computed with four different grid levels, ranging from 6 to 9, and the final time is set as $t = 378.8\ \rm{\mu s}$. The time histories of the bubble radius, obtained with different grid levels, are presented in Fig. \ref{Fig:scriven}(b). It is observed that the numerical results converge to the theoretical solution. Quantitatively, the relative errors of the bubble radius at the final time are plotted against the grid level in Fig. \ref{Fig:scriven}(d), indicating a second-order convergence rate. Furthermore, Fig. \ref{Fig:scriven}(c) presents the temperature distributions along the z-axis at the end of the simulation for different grid levels, as well as the theoretical solution. It is evident that a grid level of 8 is adequate for obtaining a converged result, which agrees well with the theoretical solution. In Fig. \ref{Fig:scriven_conservation}, the mass conservation at the highest grid level is verified by presenting the time evolutions of $\overline{M}_d$ and $\overline{M}_c$, as defined in Eqs. (\ref{Eq:md}) and (\ref{Eq:mc}). It is shown that the total mass within the domain, $\overline{M}_d$, has decreased by $15\%$, while the mass accounting for outflow, $\overline{M}_c$, has been preserved, with a maximum relative error of $0.023\%$.

To demonstrate the superiority of the ghost fluid method, we compare the velocity vectors and the superheat distributions obtained with the one fluid method and the ghost fluid method. The results at the final time obtained with different methods are shown in Fig. \ref{Fig:scrive_vel_superheat}. In the simulation with the one fluid method, the unphysical oscillations observed in the 1D tests also lead to strong spurious currents, which ultimately distort the spherical shape of the thermal boundary layer during bubble growth. In contrast, weaker spurious currents are observed with the ghost fluid method, and the spherical shape of the thermal boundary layer is well preserved. 

As depicted in Fig. \ref{Fig:stefan_problem_vel}(a), unphysical oscillations are observed solely in the cell-centered velocity $\mathbf{u}_c$, not in the face-centered velocity $\mathbf{u}_f$, when using the one fluid method. Therefore, we also performed a test where $\mathbf{u}_c$ is computed by averaging $\mathbf{u}_f$ instead of employing the approximated projection in Eq. (\ref{Eq:discretization_5}). However, as shown in Figs. \ref{Fig:scrive_vel_superheat}(c) and (d), this straightforward correction for the cell-centered velocity in the one fluid method fails to diminish the magnitude of spurious currents.

\subsection{Bubble rising in superheated liquid}
\begin{figure}[htbp]
    \centering
    \includegraphics[width=0.5\textwidth]{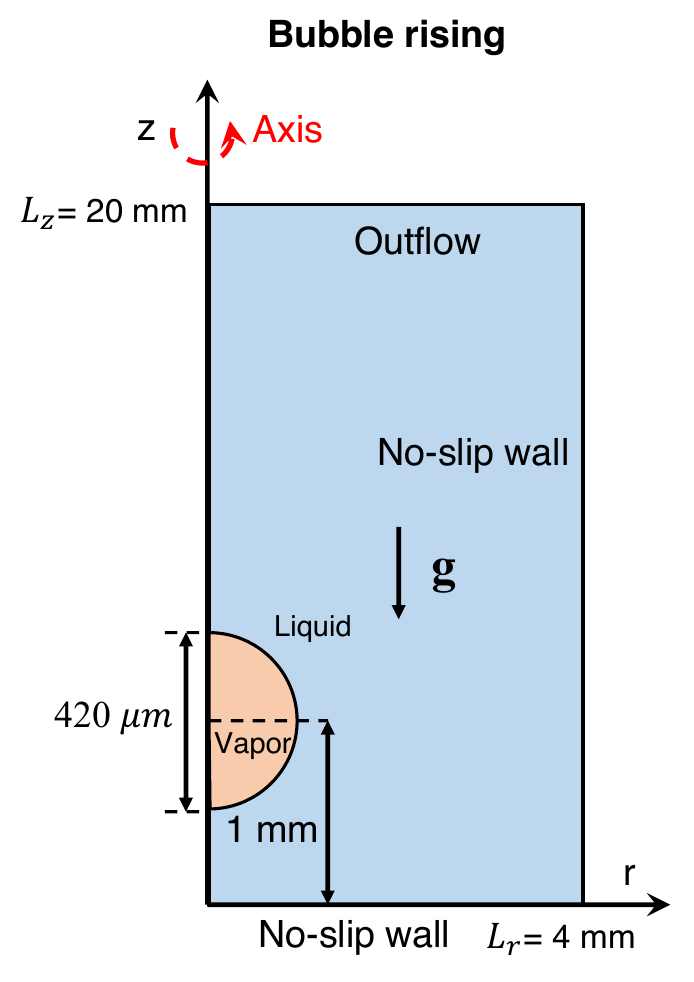}
    \caption{Schematic of the bubble rising in superheated liquid.}
    \label{Fig:schematic_bubble}
\end{figure}
\begin{figure}[htbp]
    \centering
    \includegraphics[width=1.0\textwidth]{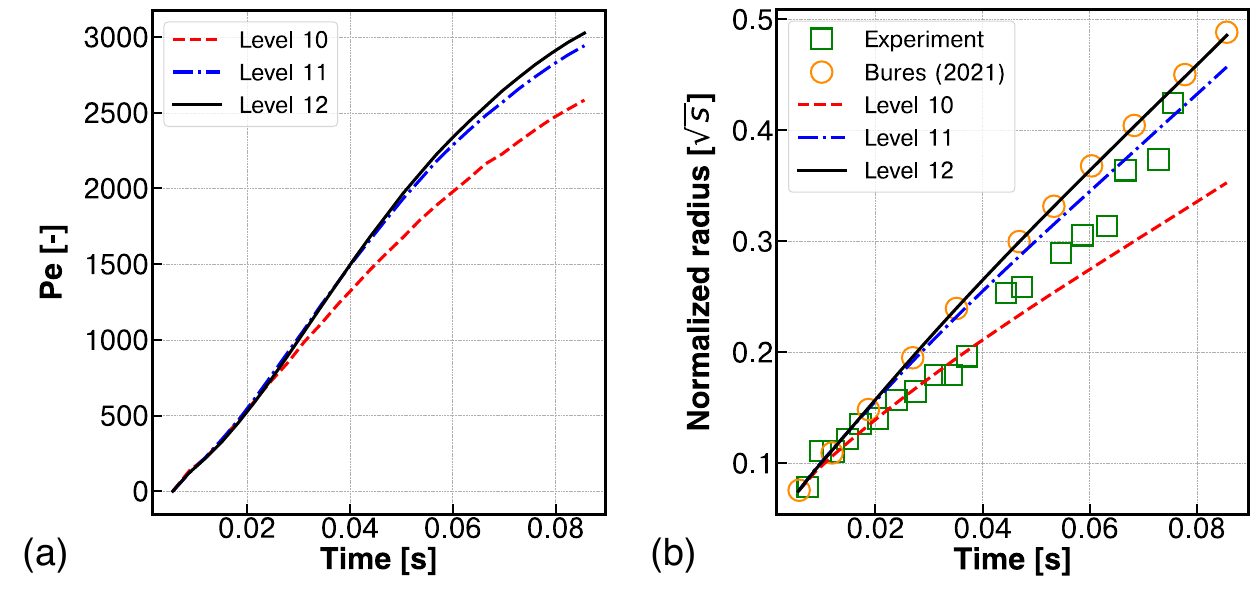}
    \caption{Bubble rising in superheated liquid: (a) The Péclet number for the bubble as a function of time obtained at different grid levels. (b) Time history of the normalized bubble radius, compared with the experimental data of Florschuetz et al. \cite{florschuetz1969growth} and the numerical result of Bure\v{s} and Sato \cite{bures2021direct}. Grid levels 10 to 12 correspond to effective grid resolutions ranging from \( 205 \times 1024 \) cells to \( 820 \times 4096 \) cells, resulting in minimum grid sizes from $19.53\ \rm{\mu m}$ to $4.88\ \rm{\mu m}$.}
    \label{Fig:bubble_rising_conv}
\end{figure}
\begin{figure}[htbp]
    \centering
    \includegraphics[width=1.0\textwidth]{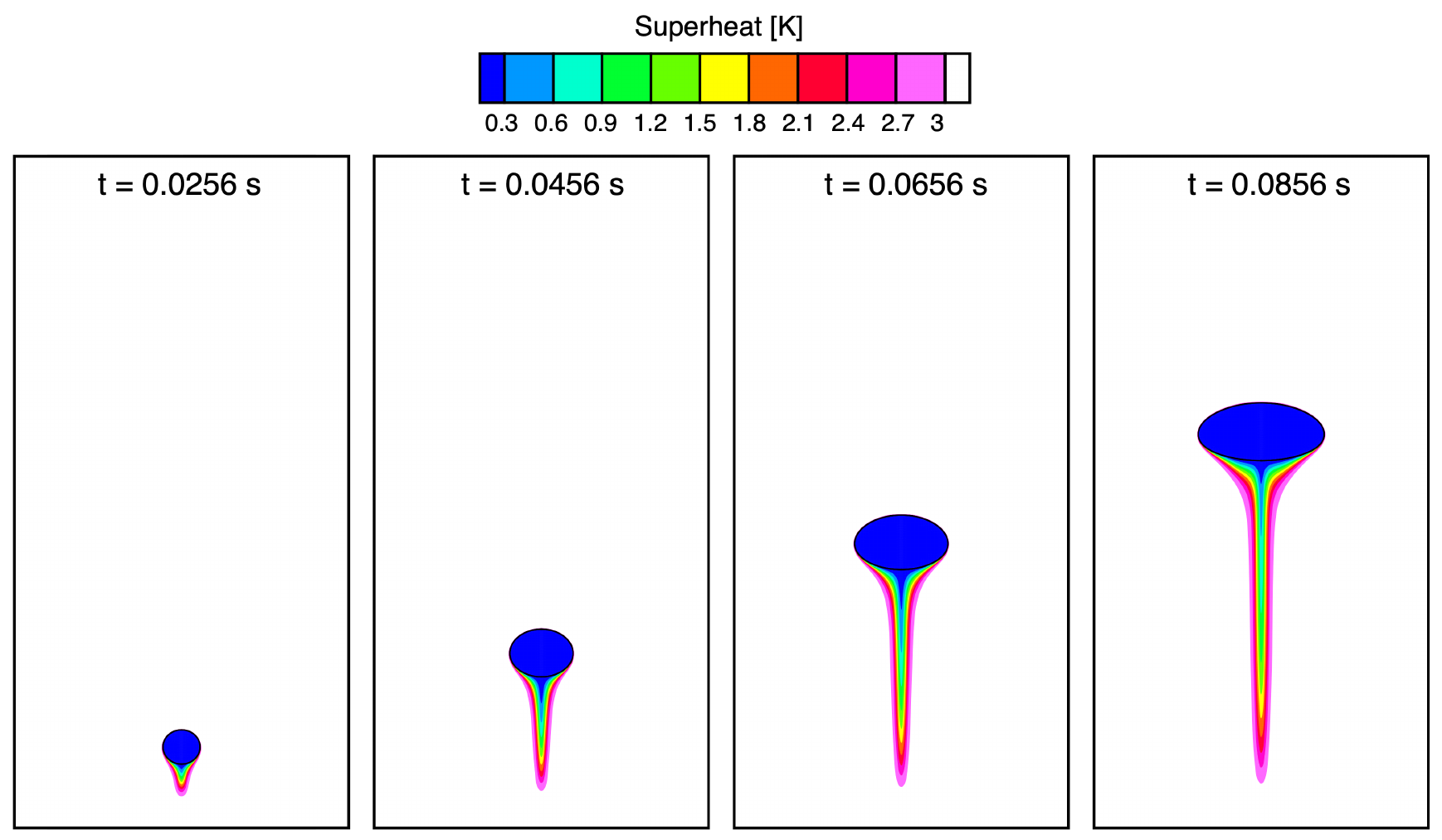}
    \caption{Bubble rising in superheated liquid: The interface shapes and superheat distributions at different time instants, obtained with grid level 12. The results are mirrored about the axis $r = 0$ for better visualization.}
    \label{Fig:bubble_rising_temp}
\end{figure}
\begin{figure}[htbp]
    \centering
    \includegraphics[width=0.8\textwidth]{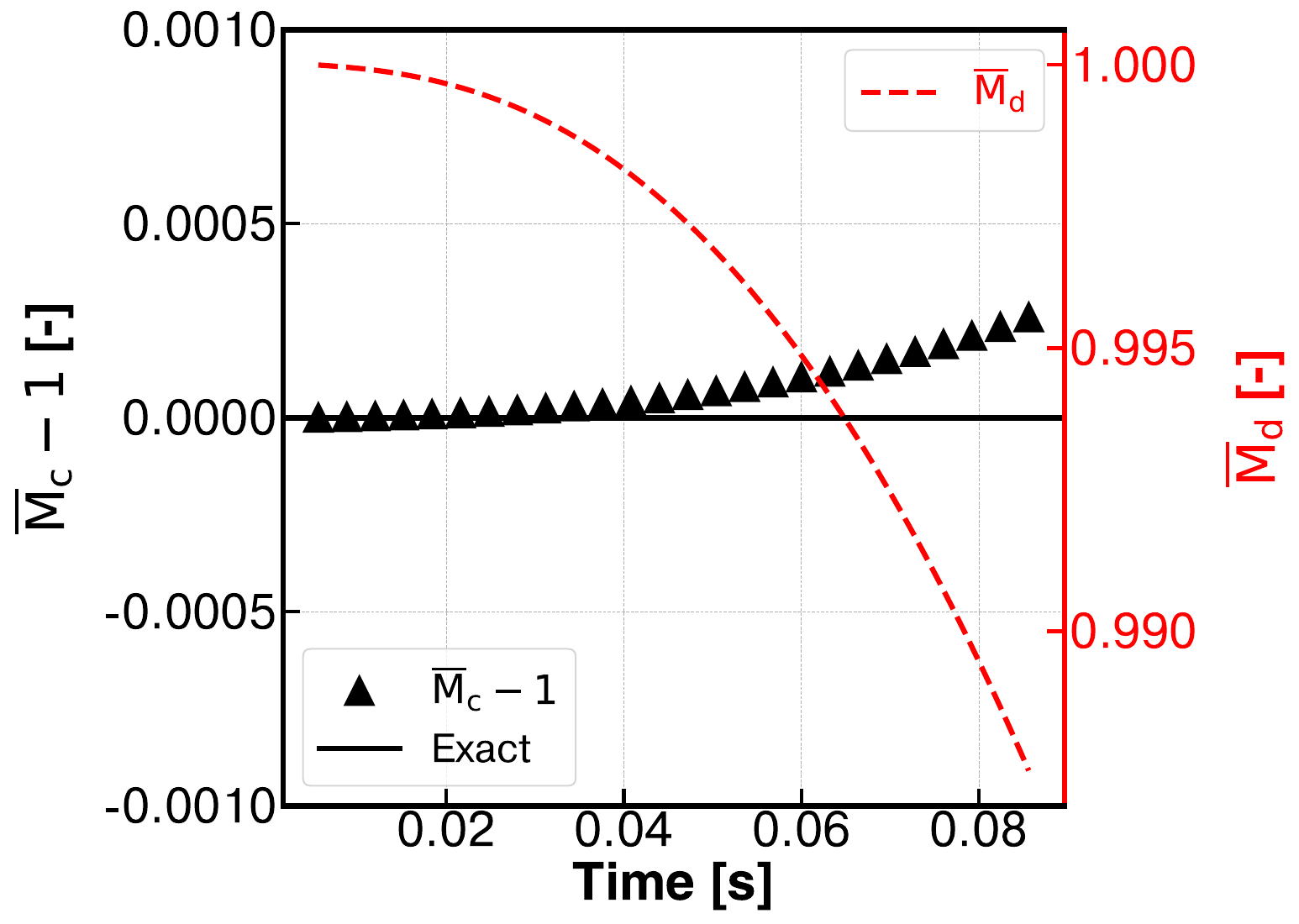}
    \caption{Relative error in mass conservation at the highest grid level for the bubble rising in superheated liquid.}
    \label{Fig:rising_conservation}
\end{figure}

Subsequently, the bubble growth in a superheated liquid under gravity is considered to validate the present method for problems involving coupled effects of phase change and buoyancy. During the rising, the bubble expands due to phase change at the interface, and deforms due to the interaction between surface tension, buoyancy, and advection effect. As a result, the bubble growth rate and the temperature distribution within the flow region cannot be described by the Scriven solution used for a spherical symmetric system \cite{tanguy2014benchmarks,sato2013sharp}. For the simulations, we follow the experiment setup of Florschuetz et al. \cite{florschuetz1969growth}, which is also simulated by Bure\v{s} and Sato \cite{sato2013sharp,bures2021direct} to verify their phase change model on axisymmetric grids. We consider an ethanol system at atmospheric pressure, with the following physical properties:
\begin{equation}
\left\{\begin{aligned}
\rho_{liq} & =757\ \mathrm{kg/m^3}, \mu_{liq}=4.29\times10^{-4}\ \mathrm{Pa\cdot s},\\ \lambda_{liq}&=0.154\ \mathrm{W/m K}, C_{p,liq} =3000\ \mathrm{J/kg\cdot K}, \\
\rho_{vap} & =1.435\ \mathrm{kg/m^3}, \mu_{vap}=1.04\times10^{-5}\ \mathrm{Pa\cdot s}, \\
\lambda_{vap} &= 0.02\ \mathrm{W/m\cdot K}, C_{p,vap} =1830\ \mathrm{J/kg\cdot K}, \\
T_{sat} & =351.45\ \mathrm{K}, T_{\infty} = 354.55\ \mathrm{K}, h_{lg} = 9.63 \times 10^5\ \mathrm{J/kg}, \sigma = 0.018 \mathrm{N/m}.
\end{aligned}\right.
\end{equation}
For the current setup, the Jakob number is $\mathrm{Ja} = 5.09$

In this case, an axisymmetric configuration is employed, with the left boundary being the axis of symmetry. As illustrated in Fig. \ref{Fig:schematic_bubble}, we use a rectangular domain with a size of $\left[ 0\ \rm{mm}, 4 \ \rm{mm} \right] \times \left[ 0\ \rm{mm}, 20 \ \rm{mm} \right]$ . The bottom and right boundaries are no-slip walls, while an outflow boundary condition is imposed on the top boundary. For the initial stage, the Scriven solution is a very good approximation as the bubble is so small that the effect of buoyancy is negligible and that its spherical shape is well preserved. Hence, initially, a bubble with a diameter of $420\ \rm{\mu m}$, which corresponds to the theoretical solution at $t = 0.0056\ \rm{s}$, is placed $1\ \rm{mm}$ above the bottom boundary. The temperature field is initialized based on the theoretical solution described by Eq. (\ref{Eq:scriven_temperature}). The gravitational acceleration, acting in the negative z-direction, is set to $9.81\ \rm{m/s^2}$.

The simulations are conducted with different grid levels, ranging from $10$ to $12$, up to $t = 0.0856\ \rm{s}$. Fig. \ref{Fig:bubble_rising_temp} shows the interface shape and superheat distribution at different time instants. As the bubble rises, its initially spherical interface deforms into an ellipsoidal shape, and a cooler region forms at the wake of the bubble. During the bubble growth, the dimensionless Péclet number can be used to compare the heat transfer resulting from convection and diffusion:
\begin{equation}
    \mathrm{Pe} = \frac{R\mathbf{u}_r}{\alpha_{l}} = \frac{\rho_{liq} C_{p,liq} R  \mathbf{u}_r }{\lambda_{liq}},
\end{equation}
where $R$ is the effective radius and $\mathbf{u}_r$ is the rising velocity, which are computed by 
\begin{equation}
R = \frac{(D_r + D_z)}{4}
\end{equation}
and
\begin{equation}
\mathbf{u}_{r} = \int_\Omega \mathbf{u} (1-f_C)d\Omega,
\end{equation}
respectively. Fig. \ref{Fig:bubble_rising_conv}(a) shows the evolution of the Péclet number as a function of time for different grid levels, confirming the convergence of our results. It can be observed that the Péclet number increases from the initial zero value to about 3000 at the end of the simulation. This trend suggests that the early stage of the process is dominated by heat diffusion, which is reasonable given that the bubble is released from rest with a small initial radius. As time progresses, the bubble grows larger and its velocity increases, leading to an increasing influence of advection on the heat transfer \cite{bures2021direct}. The time history of the normalized effective radius is compared with the experimental result of Florschuetz et al. \cite{florschuetz1969growth} and the numerical result of Bure\v{s} and Sato \cite{bures2021direct}. The normalization factor is $2\beta\sqrt{\alpha_l}$, with $\beta$ being the Scriven growth constant defined by Eq. (\ref{Eq:scriven_growth_constant}). It is shown that our results agree with the numerical solution of Bure\v{s} and Sato \cite{bures2021direct}, which was confirmed to be converged in their study. The finest grid resolution used by Bure\v{s} and Sato \cite{bures2021direct} is $6.25\ \rm{\mu m}$, while in this study, the corresponding grid resolutions for levels 10 to 12 are $19.53\ \rm{\mu m}$, $9.77\ \rm{\mu m}$, and $4.88\ \rm{\mu m}$, respectively. Generally, the numerical results are in good agreement with the experimental results, except for those obtained with grid level $10$, where the grid resolution is too coarse to resolve the thermal boundary layer accurately. A quantitative comparison of the normalized radius between the experimental data and the numerical results at different time instants is given in Table. \ref{Tab:bubble_rising_compare}. The minor discrepancy between the numerical results on the finest grid and the experimental results may be attributed to the $\pm 0.2\ \rm{K}$ uncertainties in the superheat of the liquid reported in the experimental study \cite{florschuetz1969growth}. Moreover, to check the mass conservation error, the time histories of $\overline{M}_d$ and $\overline{M}_c$, as defined in Eqs. (\ref{Eq:md}) and (\ref{Eq:mc}), are presented in Fig. \ref{Fig:rising_conservation} for the simulation at grid level 12. It is observed that $\overline{M}_d$ decreases slowly during the initial stage because of the limited area of the interface where phase change takes place. Subsequently, it decreases at an accelerating rate as the the bubble grows. The maximum relative conservation error, $\overline{M}_c - 1$, is $0.026\%$.

\begin{table}[htbp]
\centering
\caption{Bubble rising in superheated liquid: Comparison of the normalized radius between the experimental data \cite{florschuetz1969growth} and the numerical results.}
\begin{tabular}{c c c c}
\hline
 Time [s] &  Radius [$\sqrt{s}$] & Radius [$\sqrt{s}$] & Relative difference \\
 &  (experiment)  &  (present) &  \\
\hline
0.0147 & 0.122 &  0.128 & $4.69\%$\\
0.0239 &  0.157 &  0.177 & $12.74\%$\\
0.0371 &  0.197 &  0.248 & $25.89\%$\\
0.0474 &  0.256 &  0.298 & $16.41\%$\\
0.0586 &  0.306 &  0.354 & $15.69\%$\\
0.0667 &  0.364 &  0.392 & $7.97\%$\\
0.0755 &  0.425 &  0.433 & $1.88\%$\\
\hline
\end{tabular}
\label{Tab:bubble_rising_compare}
\end{table}

\subsection{Nucleate boiling in contact line regime}

\begin{figure}[htbp]
    \centering
    \includegraphics[width=0.7\textwidth]{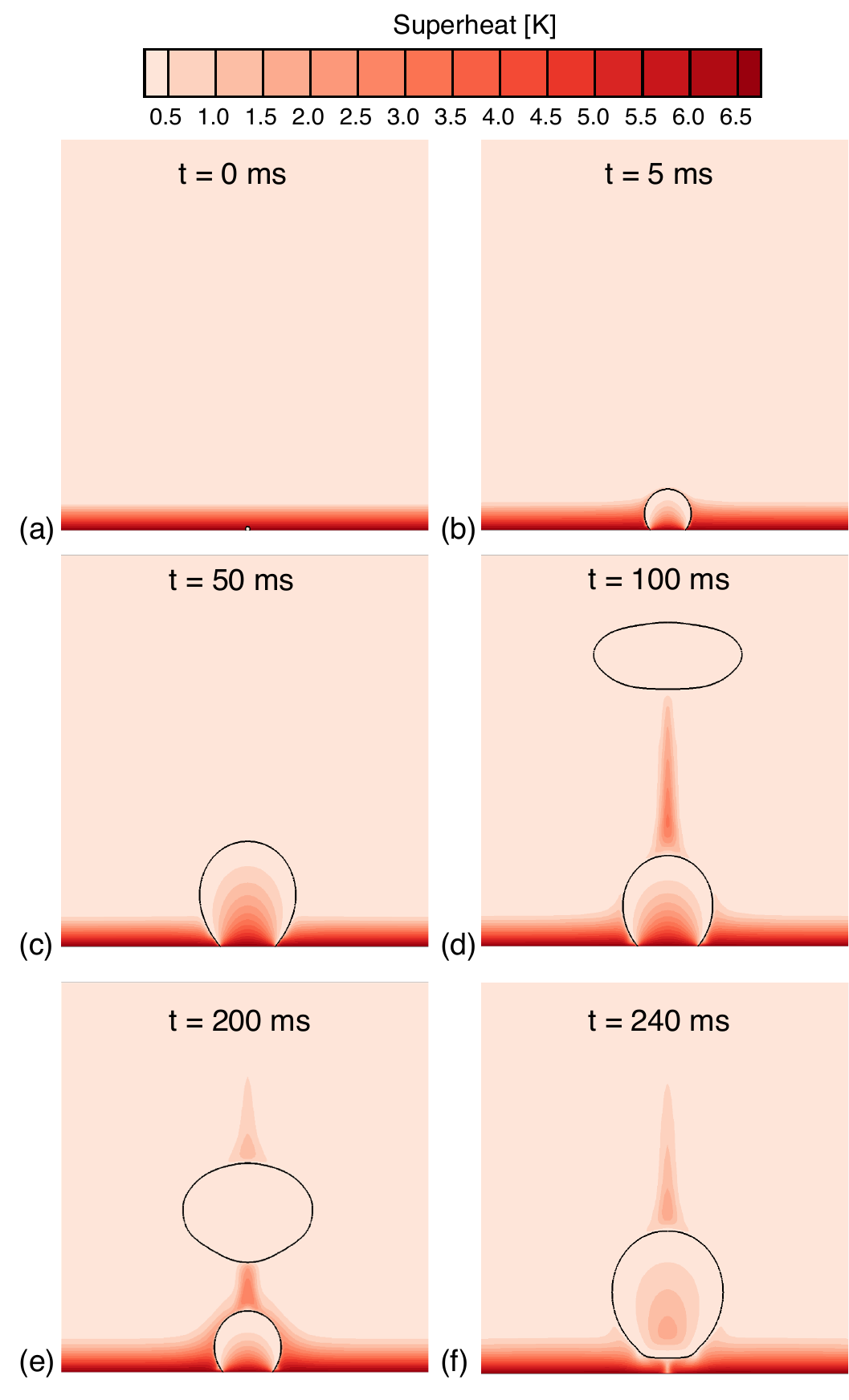}
    \caption{Nucleate Boiling: The interface shapes and superheat distributions at different time instants, obtained at grid level 11. The results are mirrored about the axis $r = 0$ for better visualization.}
    \label{Fig:nucleate_2}
\end{figure}
\begin{figure}[htbp]
    \centering
    \includegraphics[width=1.0\textwidth]{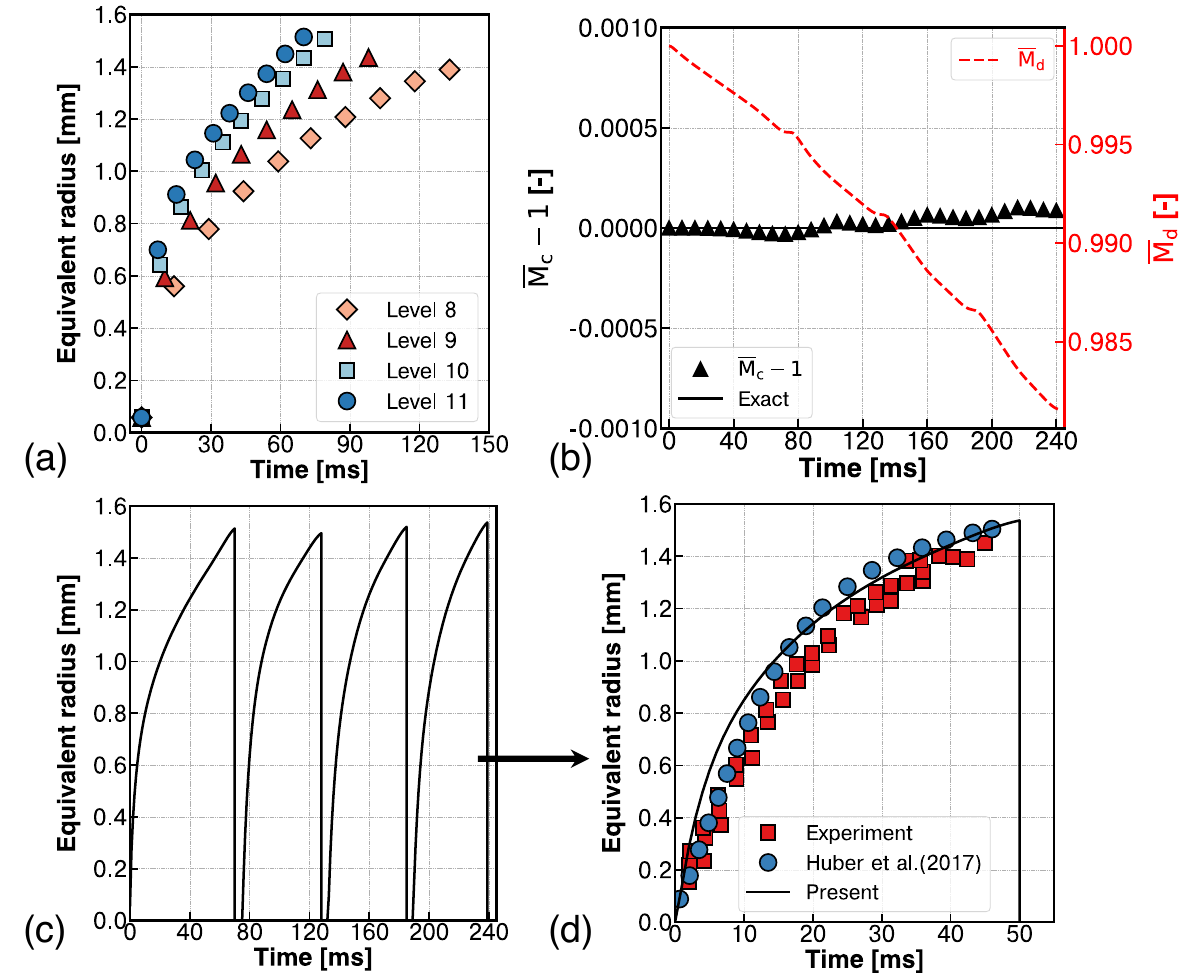}
    \caption{Nucleate Boiling: (a) Time history of the equivalent radius for the first bubble detachment at different grid levels. (b) Relative error in mass conservation. (c) Time history of the equivalent radius during the first four bubble detachment cycles. (d) Comparison of the temporal evolution of the equivalent radius for the fourth bubble detachment using the present method against previous experimental and numerical results. Grid levels 8 to 11 correspond to effective grid resolutions ranging from \( 256 \times 256 \) cells to \( 2048 \times 2048 \) cells, resulting in minimum grid sizes from $40\ \rm{\mu m}$ to $5\ \rm{\mu m}$. Note that (b)-(d) are obtained with the finest grid resolution $5\ \rm{\mu m}$, and the grid size used by Huber et al. \cite{huber2017direct} is $10\ \rm{\mu m}$. The starting time in (d) corresponds to $189\ \rm{ms}$ in (c).}
    \label{Fig:nucleate_1}
\end{figure}

Here we validate the present method for heterogeneous evaporation by simulating bubble growth on a superheated wall. Following Huber et al. \cite{huber2017direct}, we replicate the nucleate boiling experiment conducted by Son et al. \cite{son1999dynamics}. The computational domain is configured as a $10.24\ \rm{mm}$ square in an axisymmetric arrangement, with the left boundary being the axis of symmetry. Both the bottom and right boundaries are treated as no-slip walls, and an outflow boundary condition is applied at the top boundary. In this case, the movement of contact lines is involved, where a contact angle needs to be considered. In the EBIT framework, the interface curvature is computed using the height function \cite{popinet2009accurate}, with reconstructed volume fractions \cite{pan2023edge}. The contact angle can thus be imposed by taking into account the interface shapes in the boundary condition for the height function \cite{afkhami2008height}. Similar to Huber et al. \cite{huber2017direct}, conjugate heat transfer within the solid is ignored, and the temperature at the bottom wall is fixed at $T_w$. As depicted in Fig. \ref{Fig:nucleate_2}(a), the temperature field is initialized to the saturation temperature $T_{sat}$ throughout the domain, except in a thermal boundary layer near the wall where the temperature linearly varies from $T_w$ to $T_{sat}$. The thickness of this thermal boundary layer is determined using the correlation provided in Ref. \cite{kays1980convective},
\begin{equation}
   \delta_T = 7.14 \left( \frac{\mu_l \alpha_l}{g\beta_T (\Delta T)}\right)^\frac{1}{3},
\end{equation}
where $\beta_T$ represents the isobaric thermal expansion coefficient and $\Delta T$ is the wall superheat. Initially, a very small spherical nucleus with a radius of $60\ \mu\rm{m}$ is placed at the origin. It is partially truncated to match the contact angle $\theta_C$ imposed during the simulation. With $\Delta T = 7\ \rm{K}$ and $\theta_C = 50\degree$, the physical properties are chosen as
\begin{equation}
\left\{\begin{aligned}
\rho_{liq} & =958\ \mathrm{kg/m^3}, \mu_{liq}=2.82\times10^{-4}\ \mathrm{Pa\cdot s},\\ \lambda_{liq}&=0.677\ \mathrm{W/m K}, C_{p,liq} =4216\ \mathrm{J/kg\cdot K}, \\
\rho_{vap} & =0.5974\ \mathrm{kg/m^3}, \mu_{vap}=1.228\times10^{-5}\ \mathrm{Pa\cdot s}, \\
\lambda_{vap} &= 0.024\ \mathrm{W/m\cdot K}, C_{p,vap} =2034\ \mathrm{J/kg\cdot K}, \\
T_{sat} & =373.12\ \mathrm{K}, T_{w} = 380.12\ \mathrm{K}, h_{lg} = 2.256 \times 10^6\ \mathrm{J/kg}, \sigma = 0.058 \mathrm{N/m},
\end{aligned}\right.
\end{equation}
resulting in a Jakob number of $21$. Note that gravity is considered in this case, and the gravitational acceleration, acting in the negative z-direction, is set to $9.81\ \rm{m/s^2}$.

When a bubble has grown large enough, it will detach from the wall due to buoyancy. In the experiment, cold pool liquid will fill the space vacated by the departing bubble, and a new bubble will generate when the solid surface is heated to the nucleation temperature. Such a waiting period cannot be modeled in the present simulations as the conjugate heat transfer is neglected. Following Huber et al. \cite{huber2017direct}, we artificially place a new nucleus when the previous bubble moves a certain distance away from the wall, ensuring that the growth of the new bubble is not disturbed by the departed one. This distance is not reported in \cite{huber2017direct} and is chosen as $1\ \rm{mm}$ in the present study. In Figs. \ref{Fig:nucleate_2}(a)-(f), the interfaces and superheat distributions at different time instants are shown. The cycles of bubble growth and detachment can be observed. For a quantitative study, we compute the bubble equivalent radius, i.e., the radius of a spherical bubble with the same volume as the current bubble. The time evolution of the equivalent radius for the first bubble is given in Fig. \ref{Fig:nucleate_1}(a) for grid refinements from level 8 to level 11, ensuring the grid convergence. As shown in Fig. \ref{Fig:nucleate_1}(c), the problem is simulated at level 11 for four bubble detachment cycles. Following Ref. \cite{huber2017direct}, the numerical results obtained during the fourth bubble detachment cycle is compared with the experimental data. The time evolution of the equivalent radius is presented in Fig. \ref{Fig:nucleate_1}(d), from which the bubble departure radius (defined as the equivalent radius when the bubble departs) and the bubble departure frequency (defined as the inverse of the bubble growth time) can be measured. As shown in the quantitative comparison in Table. \ref{Tab:nucleate_compare}, the results of the present method are in good agreement with the experimental results and the previous numerical results reported by Huber et al. \cite{huber2017direct}. The difference between the numerical results and the experimental results can be attributed to the simplifications made in the numerical simulations, such as the neglect of conjugate heat transfer and the artificial introduction of the new nucleus. The small discrepancies between our results and those of Huber et al. \cite{huber2017direct} may be attributed to the ungiven threshold for creating new bubble nuclei. Furthermore, we have checked the mass conservation error. The time histories of $\overline{M}_d$ and $\overline{M}_c$, as defined in Eqs. (\ref{Eq:md}) and (\ref{Eq:mc}), are shown in Fig. \ref{Fig:nucleate_1}(b) for the simulation at grid level 11. It is observed that although $\overline{M}_d$ keeps decreasing due to the outflow across the top boundary, the relative conservation error $\overline{M}_c - 1$ is quite small, with the maximum value being $0.0089\%$.

\begin{table}[htbp]
\centering
\caption{Nucleate Boiling: Relative differences between the experimental data and the numerical results on bubble departure radius ($\varepsilon_r$) and bubble departure frequency ($\varepsilon_f$).}
\begin{tabular}{c c c}
\hline
 Relative difference &  Huber et al. \cite{huber2017direct} &  Present \\
\hline
 Bubble departure radius $\varepsilon_r$ &  $+5.94\%$ &  $+8.25\%$ \\
 Bubble departure frequency $\varepsilon_f$ &  $-4.95\%$ &  $-12.55\%$ \\
\hline
\end{tabular}
\label{Tab:nucleate_compare}
\end{table}

\subsection{Film boiling}
\begin{figure}[htbp]
    \centering
    \includegraphics[width=0.5\textwidth]{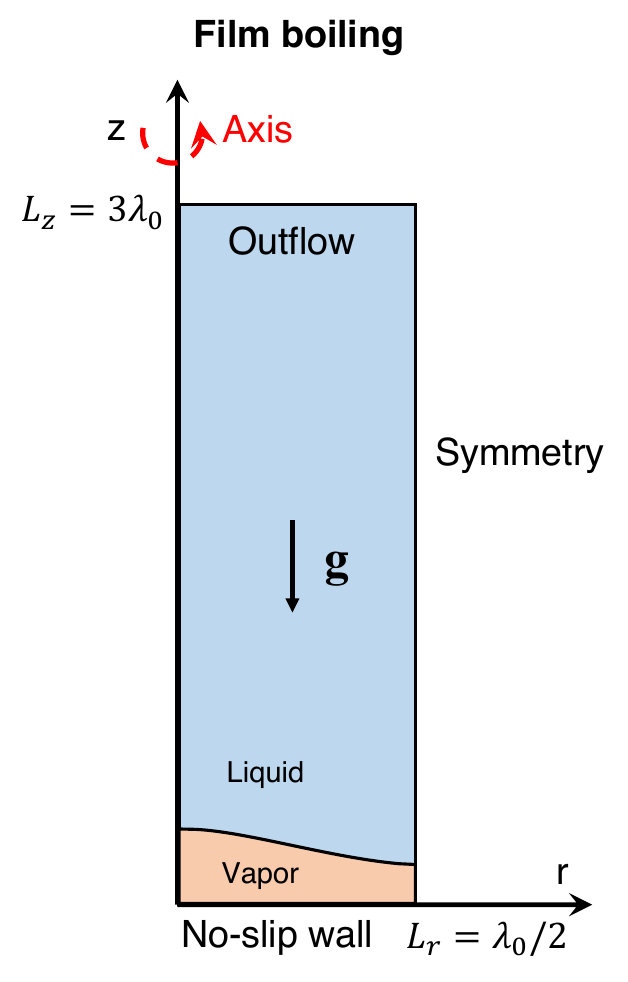}
    \caption{Schematic of the film boiling.}
    \label{Fig:schematic_film}
\end{figure}
\begin{figure}[htbp]
    \centering
    \includegraphics[width=1.0\textwidth]{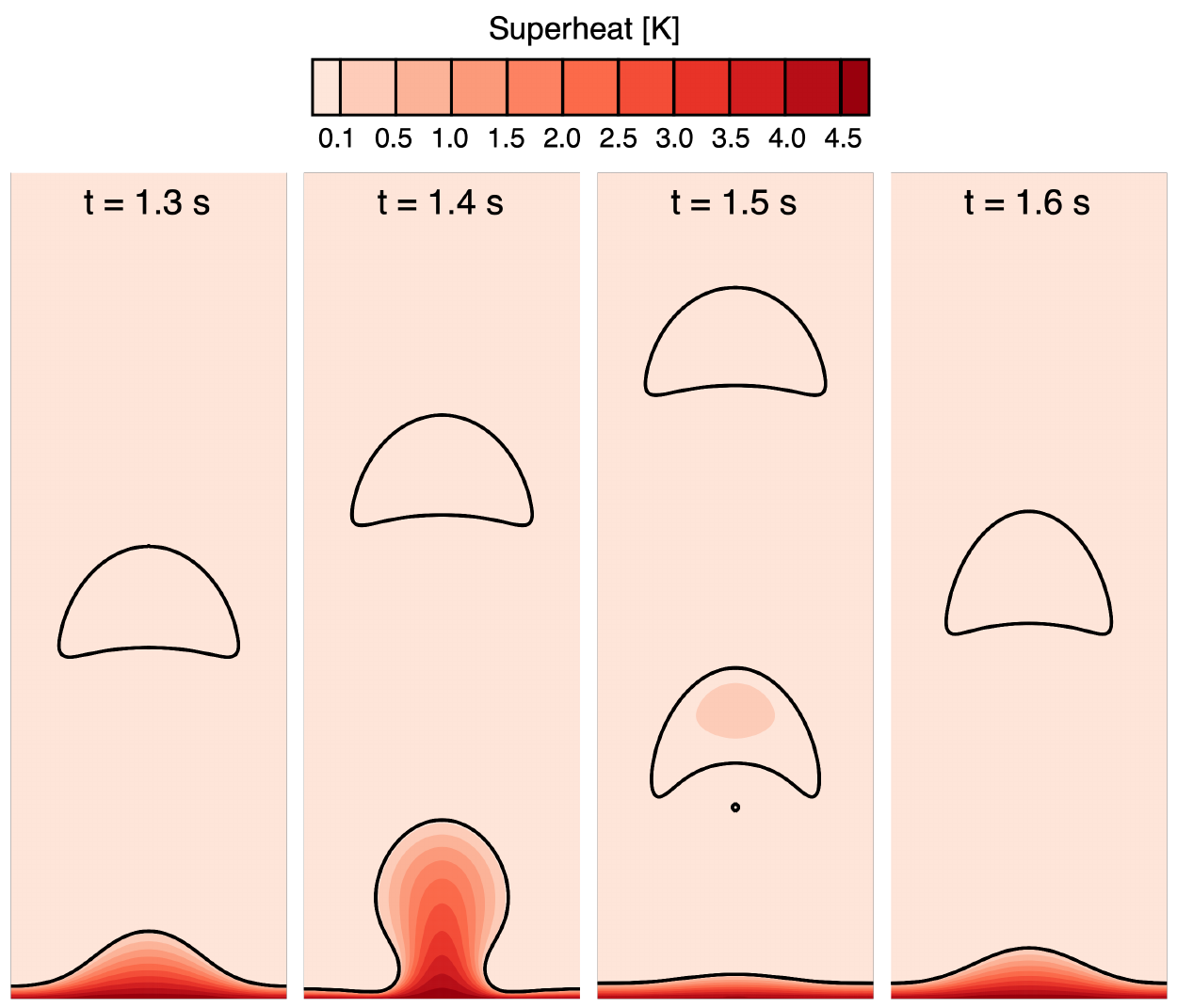}
    \caption{Film Boiling: The interface shapes and superheat distributions at different time instants, obtained with grid level 10. The results are mirrored about the axis $r = 0$ for better visualization.}
    \label{Fig:film_boiling_temp}
\end{figure}
\begin{figure}[htbp]
    \centering
    \includegraphics[width=1.0\textwidth]{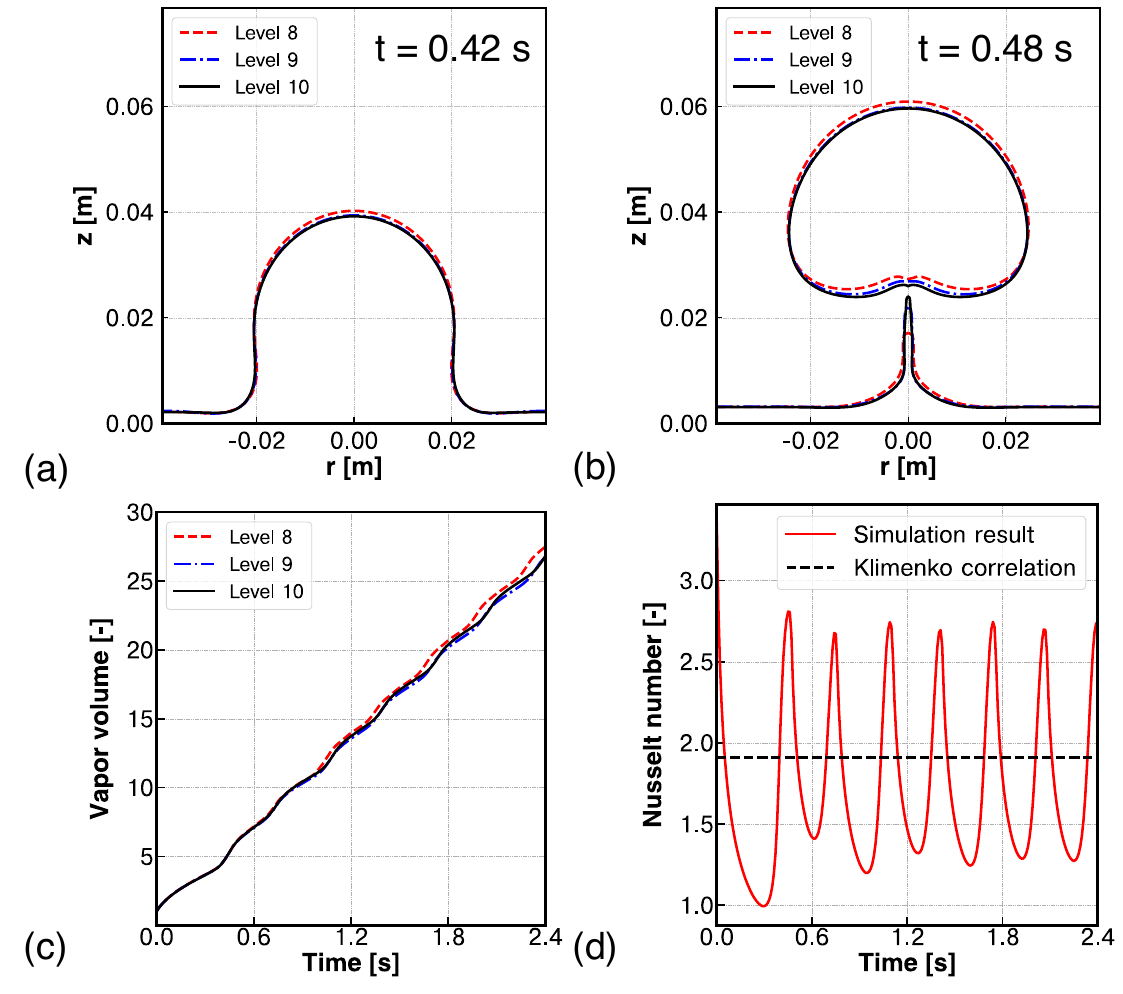}
    \caption{Film Boiling: (a) and (b) are the interface shapes obtained at different time instants with different grid levels. Note that the interfaces are mirrored about the axis $r = 0$ for better visualization. (c) The time variation of the dimensionless vapor volume, which is normalized by the initial vapor volume. (d) The comparison of the numerically computed Nusselt number at level 10 with the Klimenko correlation. Grid levels 8 to 10 correspond to effective grid resolutions ranging from \( 43 \times 256 \) cells to \( 172 \times 1024 \) cells, resulting in minimum grid sizes from $922.27\ \rm{\mu m}$ to $230.57\ \rm{\mu m}$.}
    \label{Fig:film_boiling_conv}
\end{figure}
\begin{figure}[htbp]
    \centering
    \includegraphics[width=0.8\textwidth]{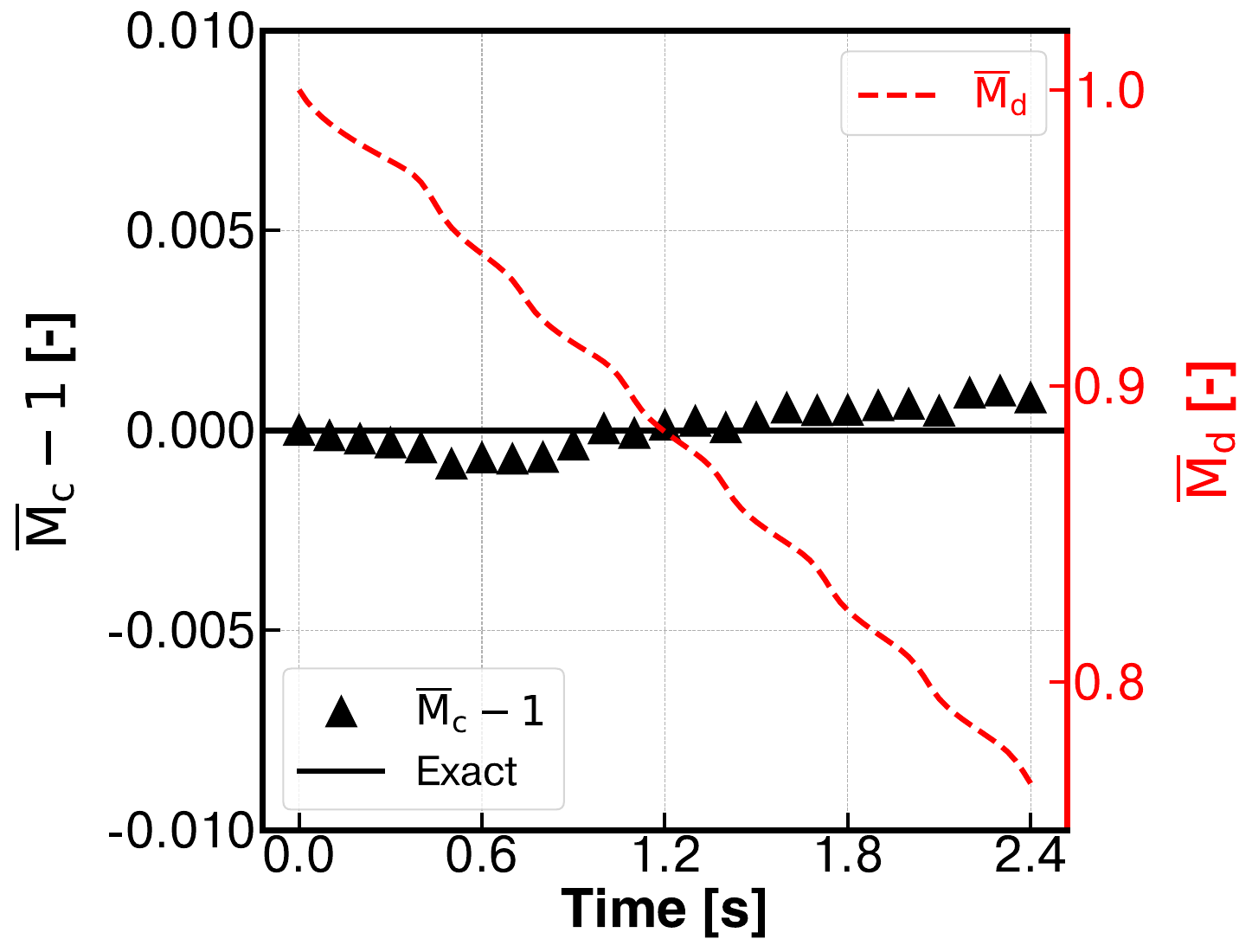}
    \caption{Relative error in mass conservation at the highest grid level for the film boiling.}
    \label{Fig:film_conservation}
\end{figure}
The next case is the film boiling problem, in which the Rayleigh–Taylor instability at the interface will be triggered by buoyancy \cite{sun2014modeling,guo2011phase}. As shown in Fig. \ref{Fig:schematic_film}(b), a superheated vapor layer is placed between a heated wall with temperature $T_w$ and a liquid at saturation temperature $T_{sat}$. An axisymmetric configuration is simulated within a domain with a size of $[0, \lambda_0/2] \times[0, 3\lambda_0]$, where $\lambda_0$ is the unstable Taylor wavelength and is calculated by
\begin{equation}
\lambda_0=2 \pi \sqrt{\frac{3 \sigma}{\left(\rho_{liq}-\rho_{vap}\right) g}}.
\end{equation}
The bottom boundary is a no-slip wall, while the outflow boundary condition and the symmetry boundary condition are applied to the top and the right boundaries, respectively. The initial interface shape is given by 
\begin{equation}
z=\frac{\lambda_0}{128}\left[4.0+\cos \left(\frac{2 \pi r}{\lambda_0}\right)\right],
\end{equation}
and the physical properties are set as
\begin{equation}
\left\{\begin{aligned}
\rho_{liq} & =200\ \mathrm{kg/m^3}, \mu_{liq}=0.1\ \mathrm{Pa\cdot s},\\ 
\lambda_{liq}&=40\ \mathrm{W/m K}, C_{p,liq} =400\ \mathrm{J/kg\cdot K}, \\
\rho_{vap} & =5\ \mathrm{kg/m^3}, \mu_{vap}=0.005\ \mathrm{Pa\cdot s}, \\
\lambda_{vap} &= 1\ \mathrm{W/m\cdot K}, C_{p,vap} =200\ \mathrm{J/kg\cdot K}, \\
T_{sat} & =1\ \mathrm{K}, T_{w} = 6\ \mathrm{K}, h_{lg} = 1 \times 10^4\ \mathrm{J/kg}, \sigma = 0.1\ \mathrm{N/m}.
\end{aligned}\right.
\end{equation}
Initially, the velocities of the two fluids are zero, and a linear temperature distribution increasing from $T_{sat}$ on the liquid-vapor interface to $T_{w}$ on the wall is imposed inside the vapor layer. The gravitational acceleration, acting in the negative z-direction, is $9.81\ \rm{m/s^2}$. Note that the current setup results in a Jakob number $\mathrm{Ja}$ of $8$ and a Taylor wavelength $\lambda_0$ of $0.0787\ \rm{m}$.

The simulations are performed up to \textcolor{blue}{$t = 2.4\ \rm{s}$}, with increasing grid levels from $8$ to $10$. Fig. \ref{Fig:film_boiling_temp} shows the interface shapes and superheat distributions at different time instants for grid level 10. It is observed that, the vapor bubble periodically forms and grows due to the boiling and Rayleigh-Taylor instability, and eventually detaches from the vapor layer. This observation is consistent with the conclusion from previous studies that film boiling is a quasi-steady phase change phenomenon \cite{sun2014modeling,guo2011phase}. The interfaces and time histories of the vapor volume, normalized to be dimensionless, are plotted in Figs. \ref{Fig:film_boiling_conv}(a), (b), and (c) for different grid levels, confirming the convergence of our results. The dimensionless vapor volume $V$ is computed as follows:
\begin{equation}
    V = \frac{\int_\Omega(1-f_C)d\Omega + V_{bc}}{V_0}, 
\end{equation}
where $V_0$ is the initial vapor volume and $V_{bc}$ represents the vapor volume flowing out from the top boundary. Additionally, we emphasize that, unlike the pure 2D case simulated in Ref. \cite{boyd2023consistent}, the bubble will pinch off in the axisymmetric configuration due to surface tension, as seen in Fig. \ref{Fig:film_boiling_conv}(b). Interested readers can refer to Ref. \cite{gibou2007level} for a detailed analysis. The quantitative study is performed using the space-average Nusselt number $\mathrm{Nu}$
\begin{equation}
\mathrm{Nu}=\frac{2}{\lambda_0}\int_0^{\lambda_0 / 2}\left(\left.\frac{\lambda^{\prime}}{T_w-T_{\mathrm{sat}}} \frac{\partial T}{\partial z}\right|_{z=0}\right) d x,
\end{equation}
where $\lambda^\prime = \sqrt{\frac{\sigma}{(\rho_{liq} - \rho_{vap})g}}$ is the characteristic length. In Fig. \ref{Fig:film_boiling_conv}(d), the time history of the Nusselt number obtained at grid level $10$ is compared with the value of $1.91$ predicted by the Klimenko correlation \cite{klimenko1981film}. The time-averaged Nusselt number from the numerical simulation is \textcolor{blue}{1.731}, indicating an \textcolor{blue}{$9.37\%$} deviation from the theoretical prediction. Furthermore, for the finest resolution, $\overline{M}_d$ and $\overline{M}_c$, as defined in Eqs. (\ref{Eq:md}) and (\ref{Eq:mc}), are plotted against time in Fig. \ref{Fig:film_conservation}. It is shown that, although $\overline{M}_d$ decreases due to the outflow, $\overline{M}_c$ is preserved around 1, with a maximum relative conservation error of $0.09\%$.

\section{Conclusion}
We have extended the Edge-Based Interface Tracking (EBIT) method \cite{chirco2023edge,pan2023edge} to the simulations of multiphase flows with phase change. By using the EBIT method, the interfacial markers are restricted to move along the grid lines, so that automatic parallelization can be achieved. Based on the EBIT method, we additionally solve the energy equations for each phase to include phase change effects. The Dirichlet boundary condition for the temperature at the interface \cite{sato2013sharp} is sharply imposed using the geometric information provided by the EBIT method. Moreover, the mass flux is determined according to the heat fluxes at the interface, which is then incorporated in the solution of the mass and momentum equations. In the presence of phase change, the velocity is not continuous across the interface, leading to numerical instability. This issue is worse on the collocated grids as the cell-centered velocity is approximately projected \cite{popinet2009accurate}. Instead of solving a second Poisson equation \cite{zhao2022boiling} to suppress such oscillations, we have shown that this issue can be addressed by using the ghost fluid method \cite{tanguy2014benchmarks}. In the ghost fluid method, two velocities are employed for each phase, and the coupling between two phases is achieved by populating the ghost velocities. As the ghost velocity is set according to the jump condition, the discontinuity at the interface is removed, thus the numerical stability of the present method is improved. Several benchmark problems have been simulated to validate the present method. It is shown that the present method agrees very well with the theoretical solutions and the experimental results. The present method has been implemented in the free software Basilisk \cite{popinet2015quadtree}. The developed codes, along with the configurations for all tests, are freely available in the Basilisk sandbox \cite{tiansandbox}. 

In multiscale simulations, Front-Tracking methods can offer several advantages over Front-Capturing methods. For instance, FT allows for handling thin strips of fluid that are smaller than the grid size, a task that proves challenging with VOF or LS. Another often-used multiscale feature in both VOF and LS is the conversion of small connected patches into Lagrangian Particles. This process is facilitated with FT, as the exact location of the interface is known, and its topology is determined. Moreover, as FT tracks the interface without diffusion and accurately provides the exact position of the interface, they are very suitable for coupling with multiscale models on the interface. For instance, in Ref. \cite{aboulhasanzadeh2012multiscale}, a mass boundary layer model is employed on the interface of buoyant bubbles, aiming to improve computational efficiency for mass transfer problems characterized by high Schmidt number. However, the parallelization for traditional Front-Tracking methods is challenging. In contrast, the EBIT method allows for automatic parallelization with quad/octree AMR, which is beneficial for studying flows with a wide range of scales. Thus, we believe that the EBIT method can be extended appropriately and bring progress in the multiscale simulations of phase change flows. Several extensions are under consideration and under development, such as the 3D EBIT method, the EBIT method allowing for two markers per cell edge, and the coupling between the DNS solver and subgrid models effectively resolving the thin thermal boundary layer at high Jakob numbers. Applying the present method to more studies of nucleate boiling is also one of our main goals.

\section{Acknowledgements}
This project has received funding from the European Research Council (ERC) under the European Union’s Horizon 2020 research and innovation programme (grant agreement number 883849). We are grateful to GENCI for generous allocation on Adastra supercomputers.



\bibliographystyle{model1-num-names}
\bibliography{ref}
\end{document}